\documentclass[12pt]{iopart}

\usepackage{mathtext}
\usepackage{graphicx}
\usepackage{color}

\newcommand{\vvec}[1]{\mathbf{#1}}
\newcommand{\mathbb}[1]{\mathbf{#1}}
\newcommand{\idx}[1]{_{\mathrm{#1}}}

\begin{document}

\title[Universal shape characteristics]{Universal shape characteristics for the mesoscopic polymer chain \textit{via} dissipative particle dynamics}

\author{O.~Kalyuzhnyi$^{1,4}$, J.M.~Ilnytskyi$^{1,4}$, Yu.Holovatch$^{1,4}$}
\author{C. von Ferber$^{2,3,4}$}

\address{$^{1}$Institute for Condensed Matter Physics, National Acad. Sci. of Ukraine, UA--79011 Lviv, Ukraine}
\address{$^{2}$Applied Mathematics Research Centre, Coventry University, Coventry, CV1 5FB, United Kingdom}
\address{$^{3}$Heinrich-Heine Universit\"at D\"usseldorf, D-40225 D\"usseldorf, Germany}
\address{$^{4}$Doctoral College for the Statistical Physics of Complex Systems, Leipzig-Lorraine-Lviv-Coventry
              $({\mathbb L}^4)$, D-04009 Leipzig, Germany}

\begin{abstract}
In this paper we study the shape characteristics of a polymer chain
in a good solvent using a mesoscopic level of modelling. The
dissipative particle dynamics simulations are performed in the $3D$
space at a range of chain lengths $N$. The scaling laws for the
end-to-end distance and gyration radius are examined first and found
to hold for $N\geq 10$ yielding reasonably accurate value for the
Flory exponent $\nu$. Within the same interval of chain lengths, the
asphericity, prolateness, size ratio and other shape characteristics
of the chain are found to become independent of $N$. Their mean
values are found to agree reasonably well with the respective
theoretical results and lattice Monte Carlo simulations. Broad
probability distributions for the shape characteristics are found
resembling in form the results of lattice Monte Carlo simulations.
By means of analytic fitting of these distributions the most
probable values for the shape characteristics are found to
supplement their mean values.
\end{abstract}

%
\vspace{2pc}

%

\date{\today}

\maketitle

\section{\label{I}Introduction}

Universal scaling laws for the dimensional properties of flexible
polymer chain in a good solvent, such as the end-to-end distance
$R_e$ and the gyration radius $R_g$, are well explained and
understood since the groundbreaking studies by de Gennes and des
Cloizeaux \cite{deGennes1979,desCloizeaux1982}. It was found that
both properties scale as
\begin{equation}\label{Re_Rg_scl}
\langle R^2_e \rangle \sim \langle R^2_g \rangle \sim N^{2\nu}
\end{equation}
for large enough numbers of monomers $N$, where the exponent $\nu$
is universal and depends on the dimension of space $d$ only. Besides
that, the probability distribution $p(R_e)$ is also examined in both
asymoptotic regimes of small and large values of  $R_e$ based on the
number of chain conformations \cite{deGennes1979,desCloizeaux1982}.
Similar arguments allowed Lhuillier to suggest an heuristic form for
the probability distribution $p(R_g)$ as well \cite{Lhuillier1988}.

As far as the scaling laws are also valid for a number of other
characteristics of the polymer chain, the universality and scaling
properties are often treated as identical concepts. This, however,
is a misconception as far as there exist characteristics that
are universal but do not obey scaling laws. An example is given by
the shape characteristics of polymer chains that which is in the
focus of this paper. These are important in a number of
applications, to mention here various catalytic activities
\cite{Arunchander2015,Anderson2016} and gel chromatography
\cite{Striegel2009}.

The fact that the shape of a polymer coil  in a good solvent is not
spherical is known since a classical work by Kuhn \cite{Kuhn1936}.
However, the understanding that certain shape properties of polymer
chains are universal and depend, like scaling parameters, solely on
$d$, is brought \textit{via} the use of the renormalization group
method (see, e.g. \cite{Guida1998,ZinnJustin2002}). These findings
were also supported by a number of  numerical simulations performed
on lattice models of the self avoiding walk (SAW)
\cite{Aronovitz1986,Cannon1991,Jagodzinski1992,Bishop1988,Benhamou1985,Diehl1989,Blavatska2011,Zifferer1999a,Zifferer1999b}.
One should note that the mean values for most shape characteristics
are different but nonetheless close to their random walk (RW)
counterparts. Some important implication here can be seen in the
fact that the probability distributions for most of these
characteristics are very broad and asymmetric yielding an ambiguity
in the definition of their mean values. The shape of these
distributions remains to be explained in a manner similar to the one
given for their counterparts $p(R_e)$
\cite{deGennes1979,desCloizeaux1982} and
$p(R_g)$\cite{Lhuillier1988,Victor1990}.

Most of the simulation studies mentioned above use the Monte Carlo
algorithm applied to the SAW lattice model. This approach achieves
very good configuration statistics by means of relatively low
computation cost. Atomistic off-lattice models, on the other hand,
allow one to include the effects of chain stiffness, chain
composition, the role of solvent, etc. in a chemical way, but at an
increased simulation time cost. A good compromise here is the use of
coarse-grained approaches that combine the best of two worlds:
chemical versatility and computational efficiency. One of such
approaches is the dissipative particle dynamics (DPD) method
\cite{Hoogerbrugge1992,Espanol1995}, that has already been used by a
number of authors
\cite{Schlijper1995,Kong1997,Spenley2000,Symeonidis2005,Jiang2007,Nardai2009}
including two of the current authors \cite{Ilnytskyi2007}, to
examine the scaling properties of a polymer chain in a solvent of
variable quality.

The benefit of this approach is that it provides the means to study
a number of important problems related to microphase separation of
amphi- and polyphilic molecules, self-assembly, adsorption, etc.
(see, e.g.
\cite{Ilnytskyi2008,Ilnytskyi2011,Ilnytskyi2013,Ilnytskyi2016}).
Macromolecular shape plays an important role in all of the above
mentioned problems. Therefore, the validation of this approach with
respect to its reliability in predicting the correct macromolecular
shape is of high practical interest.

This is exactly at the heart of the current study, were we
apply DPD simulations to the simplest case -- a single chain in a
good solvent. We examine the asphericity, prolateness and a number
of the other shape characteristics, as well as their probability
distributions. Besides the mean values for each of these properties,
we also evaluate their most probable values by means of fitting
their probability distributions to respective analytic expressions.
The outline of the study is as follows. The simulation approach and
the properties of interest are described in Sec.~\ref{II}, scaling
properties and a probability distributions analysis are covered in
Sec.~\ref{III}, probability distributions for shape characteristics
are discussed in Sec.~\ref{IV} and conclusions are provided in
Sec.~\ref{V}.

\section{\label{II}Simulation approach and properties of interest}

Our study is based on mesoscopic DPD simulations, which has two main
advantages: (i) neglecting less important properties of a system on
small length- and time-scales, and (ii) preserving hydrodynamic
limits \cite{Hoogerbrugge1992,Espanol1995}. A single polymer chain
and an explicit solvent are contained within a cubic simulation box
with a linear size of at least $5R'_g$. Here, $R'_g$ is the estimate
for the gyration radius of a chain of $N$ monomers in a good
solvent: $R'_g\approx [b(N-1)]^{0.59}$ ($b\approx 0.9$ is an
estimate for an average bond length as reported in
\cite{Ilnytskyi2007}). The monomers are soft beads of equal size,
and each one represents either a fragment of a real polymer chain or
a few molecules of a solvent. We restrict our study to the case of
an athermal solvent, where all types of pairwise bead-bead
interactions: polymer-polymer, solvent-solvent and polymer-solvent
are identical.

The monomer coordinates $\vvec{x}_i$ are defined in continuous space
(in contrary to most studies using the Monte Carlo method for
similar studies, which, typically, are performed on a lattice). This
allows to describe the phase space of conformations by chains of
shorter length. On the other hand, the soft nature of DPD
interaction prevents the system from being caught in a
metastable state (what is often observed in the case of the
molecular dynamics when atomistic potentials are applied).

We follow the DPD approach as described in Ref.~\cite{Groot1997}.
The length is represented in units of the diameter of the soft bead,
and the energy scale is assumed to be $\epsilon^*=k_{B}T=1$, where
$k_{B}$ is the Boltzmann constant, $T$ is the temperature and time
is expressed in $t^*=1$. The monomers are connected \textit{via}
harmonic springs, which results in a force
\begin{equation}\label{FB}
  \vvec{F}^B_{ij} = -k\vvec{x}_{ij}\,,
\end{equation}
where $\vvec{x}_{ij}=\vvec{x}_i-\vvec{x}_j$ and $k$ is the spring
constant. The non-bonded forces contain three contributions
\begin{equation}
  \vvec{F}_{ij} = \vvec{F}^{\mathrm{C}}_{ij} + \vvec{F}^{\mathrm{D}}_{ij}
  + \vvec{F}^{\mathrm{R}}_{ij}\,,
\end{equation}
where $\vvec{F}^{\mathrm{C}}_{ij}$ is the conservative force,
resulting from the repulsion between $i$-th and $j$-th soft beads,
$\vvec{F}^{\mathrm{D}}_{ij}$ is the dissipative force, that occurs
due to the friction between soft beads, and random force
$\vvec{F}^{\mathrm{R}}_{ij}$ that works in pair with a dissipative
force to thermostat the system. The expressions for all these three
contributions are given below \cite{Groot1997}
\begin{equation}\label{FC}
  \vvec{F}^{\mathrm{C}}_{ij} =
     \left\{
     \begin{array}{ll}
        a(1-x_{ij})\displaystyle\frac{\vvec{x}_{ij}}{x_{ij}}, & x_{ij}<1,\\
        0,                       & x_{ij}\geq 1,
     \end{array}
     \right.
\end{equation}
\begin{equation}\label{FD}
  \vvec{F}^{\mathrm{D}}_{ij} = -\gamma
  w^{\mathrm{D}}(x_{ij})(\vvec{x}_{ij}\cdot\vvec{v}_{ij})\frac{\vvec{x}_{ij}}{x^2_{ij}},
\end{equation}
\begin{equation}\label{FR}
  \vvec{F}^{\mathrm{R}}_{ij} = \sigma
  w^{\mathrm{R}}(x_{ij})\theta_{ij}\Delta t^{-1/2}\frac{\vvec{x}_{ij}}{x_{ij}},
\end{equation}
where $x_{ij}=|\vvec{x}_{ij}|$,
$\vvec{v_{ij}}=\vvec{v}_{i}-\vvec{v}_{j}$, $\vvec{v}_{i}$ is the
velocity of $i$th bead, $a$ is the amplitude for the conservative
repulsive force. The dissipative force has an amplitude $\gamma$ and
decays with the distance according to the weight function
$w^{\mathrm{D}}(x_{ij})$. The amplitude for the random force is
$\sigma$ and the respective weight function is
$w^{\mathrm{R}}(x_{ij})$. $\theta_{ij}$ is the Gaussian random
variable and $\Delta t$ is the time-step of the simulations. As was
shown by Espa\~{n}ol and Warren \cite{Espanol1995}, to satisfy the
detailed balance requirement, the amplitudes and weight functions
for the dissipative and random forces should be interrelated:
$\sigma^2=2\gamma$ and
$w^{\mathrm{D}}(x_{ij})=\large[w^{\mathrm{R}}(x_{ij})\large]^2$.

The system is soft repulsive and is kept together by an external
pressure to provide the required density, which corresponds to a
liquid state. In these simulations the following numeric values are
used: $a=25$ for all pairs of interacting beads, $\gamma=6.75$,
$\sigma=\sqrt{2\gamma}=3.67$ and the time-step $\Delta t=0.04$. The
duration of all runs (each performed for a different chain length
$N=5-40$) was fixed at $8\cdot 10^6$ DPD steps.

Here and thereafter we consider the case of space dimension $d=3$
only. All shape characteristics of a chain are derived from the
components of the instantaneous gyration tensor $\vvec{Q}$ defined
as in \cite{Solc1971a,Solc1971b}:
\begin{equation}\label{Q_def}
Q_{\alpha\beta} = \frac{1}{N}
\sum_{n=1}^{N}(x^{\alpha}_{n}-X^{\alpha})(x^{\beta}_{n}-X^{\beta})
\hspace{1em} \alpha,\beta=1,2,3.
\end{equation}
Here, $N$ is the number of monomers of a chain, $x_n^{\alpha}$
denotes the set of  the Cartesian coordinates of $n$th monomer
center: $\vvec{x}_n=(x_n^1,x_n^2,x_n^3)$, and $X^{\alpha} =
\frac{1}{N}\sum_{n=1}^N x_{n}^{\alpha}$ are the coordinates of the
center of mass for the chain. Its eigenvectors define the axes of a
local frame of a chain and the mass distribution of the latter along
each axis is given by the respective eigenvalue $\lambda_i$,
$i=1,2,3$, respectively. The trace of $\vvec{Q}$ is an invariant
with respect to rotations and is equal to an instantaneous squared
gyration radius of the chain
\begin{equation}\label{Rg_def}
R_g^2 = \mathrm{Tr}\,\vvec{Q} = 3\bar{\lambda}.
\end{equation}
Here, the average over three eigenvalues, $\bar{\lambda}$, is
introduced to simplify the following expressions.

The instantaneous asphericity $A$ (sometimes also referred to as
``the relative shape anisotropy''). The prolateness $S$ and size
ratio $g$ are defined as
\cite{Aronovitz1986,Zifferer1999a,Zifferer1999b,Blavatska2011}
\begin{equation}\label{ASg_def}
A = \frac{1}{6}\frac{\sum_{i=1}^{3}(\lambda_{i}-\bar{\lambda})^{2}}{\bar{\lambda}^{2}},\;\;\;
S = \frac{\prod^{3}_{i=3}(\lambda_{i}-\bar{\lambda})}{\bar{\lambda}^{3}},\;\;\;
g = \frac{R_e^2}{R_g^2},
\end{equation}
where $R_e$ is the magnitude of the end-to-end vector
$\vvec{R}_e=\vvec{x}_N-\vvec{x}_1$. A spherical shape is
characterised by $A=S=0$, whereas for a non-spherical one it is:
$0<A<1$ and $0<S<2$ (for prolate shape) and $-1/4<S<0$ (for oblate
shapes). To define the remaining shape characteristics we follow
Ref.~\cite{Blavatska2011}. To this end we introduce the
following triplet of vectors: $\vvec{r}_1$, $\vvec{r}_2$ and
$\vvec{r}_3$. Here $\vvec{r}_1\equiv\vvec{R}_e$, $\vvec{r}_2$ is the
component of the vector $\vvec{x}_{N/2}-\vvec{x}_1$ perpendicular to
$\vvec{r}_1$, whereas $\vvec{r}_3$ is the component of the vector
$\vvec{x}_{N/4}-\vvec{x}_1$ perpendicular to both $\vvec{r}_1$ and
$\vvec{r}_2$. For more details, see e.q. Fig.~1 in
Ref.~\cite{Blavatska2011}. On a technical note, for the case of a
non-integer $N/2$ (or $N/4$), the vector $\vvec{x}_{N/2}$ (or
$\vvec{x}_{N/4}$) was chosen \textit{via} interpolation between the
vectors $\vvec{x}_{n}$ and $\vvec{x}_{n+1}$ for the adjacent
monomers, where $n=\mathrm{int}(N/2)$ [or $n=\mathrm{int}(N/4)$].
The magnitudes of $\vvec{r}_1$, $\vvec{r}_2$ and $\vvec{r}_3$ are
denoted as $r_1$, $r_2$ and $r_3$, respectively. Then, the ratios
\begin{equation}\label{r12_r13_def}
r_{12} = \frac{r_1}{r_2},\;\;\;r_{13} = \frac{r_1}{r_3}
\end{equation}
are evaluated at each time instance.

In real experiments one observes shape characteristics of polymer
chains averaged over a sample and over time trajectory. The same can
be done in the course of the DPD simulation study. We will  denote
this hereafter as $\langle ~\cdots~ \rangle$. Let us note that
all expressions [Eqs.~ (\ref{ASg_def}) -- (\ref{r12_r13_def})]
contain these ratios. Therefore, the averaging can be performed in
two ways: as a ratio of the averages or, alternatively, as the
average of the ratios, both ways are used in the literature
\cite{Aronovitz1986,Jagodzinski1992,Zifferer1999a,Zifferer1999b,Blavatska2011}.
The former definition leads to the first set of the averages:
\begin{equation}\label{ASg_hat}
\hat{A} =
\frac{1}{6}\frac{\langle\sum_{i=1}^{3}(\lambda_{i}-\bar{\lambda})^{2}\rangle}{\langle\bar{\lambda}^{2}\rangle},\;\;\;
\hat{S} =
\frac{\langle\prod^{3}_{i=1}(\lambda_{i}-\bar{\lambda})\rangle}{\langle\bar{\lambda}^{3}\rangle},\;\;\;
\hat{g} = \frac{\langle R_e^2\rangle}{\langle R_g^2\rangle},
\end{equation}
\begin{equation}\label{r12_r13_hat}
\hat{r}_{12} = \frac{\langle r_1\rangle}{\langle r_2\rangle},\;\;\;
\hat{r}_{13} = \frac{\langle r_1\rangle}{\langle r_3\rangle},
\end{equation}
%
whereas the latter definition yields a second set of averages:
\begin{equation}\label{ASg_av}
\langle A\rangle =
\frac{1}{6}\left\langle\frac{\sum_{i=1}^{3}(\lambda_{i}-\bar{\lambda})^{2}}{\bar{\lambda}^{2}}\right\rangle,\;\;
\langle S\rangle =
\left\langle\frac{\prod^{3}_{i=1}(\lambda_{i}-\bar{\lambda})}{\bar{\lambda}^{3}}\right\rangle,\;\;
\langle g\rangle = \left\langle\frac{R_e^2}{R_g^2}\right\rangle,
\end{equation}
\begin{equation}\label{r12_r13_av}
\langle r_{12}\rangle =
\left\langle\frac{r_1}{r_2}\right\rangle,\;\;\; \langle
r_{13}\rangle = \left\langle\frac{r_1}{r_3}\right\rangle.
\end{equation}
%

\section{Shape characteristic: scaling and the mean values}\label{III}

%
\begin{figure}[!h]
\centering
\includegraphics[width=8cm,angle=270]{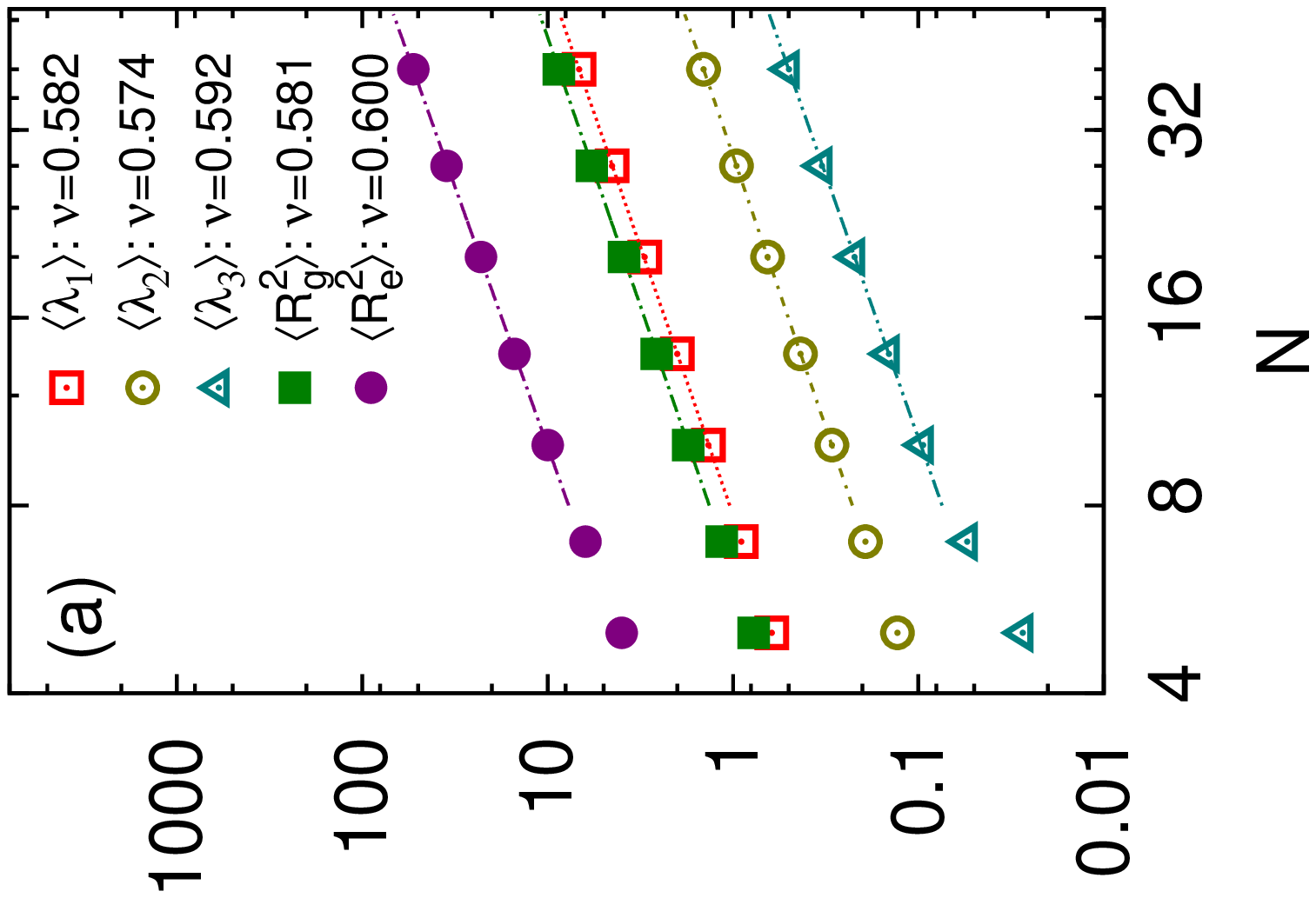}\hspace{-3em}
\includegraphics[width=8cm,angle=270]{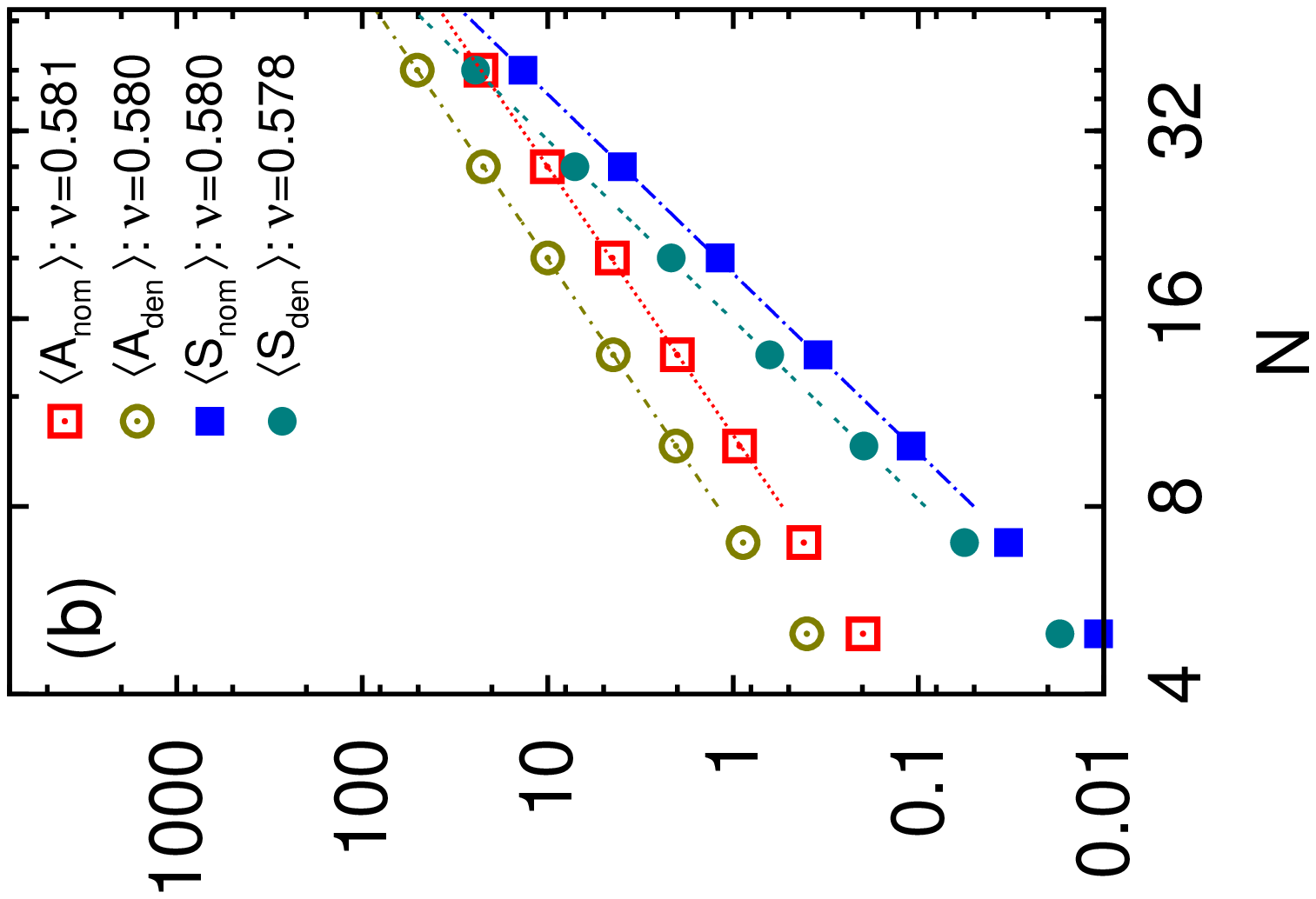}
\caption{\label{scaling_L123Rgsq}(a) Log-log plot for the average
eigenvalues $\langle \lambda_{\alpha}\rangle$ of the gyration
tensor, the squared gyration radius $\langle R_g^2\rangle$ and the
squared end-to-end distance $\langle R_e^2\rangle$. (b) the same for
the average values of asphericity and prolateness nominators and
denominators as defined in Eq.~(\ref{AS_nom_den}). $N$ is the number
of monomers per chain. Note, that different angles, observed for the
two sets of curves in this figure are due to the fact that scaling
of $A\idx{nom}$ and $A\idx{den}$ is governed by the exponent 4$\nu$,
whereas that of $S\idx{nom}$ and $S\idx{den}$ -- by $6\nu$, see
Eq.~(\ref{nom_den_scl}).}
\end{figure}
The scaling laws (\ref{Re_Rg_scl}) for the polymer chain in a good
solvent modelled by means of the DPD method have previously been
discussed in detail \cite{Ilnytskyi2007}. Therefore, we
will recall these rather briefly here. In particular, as indicated
in Fig.~\ref{scaling_L123Rgsq} (a), at $N\geq 10$ both $\langle
R_g^2\rangle$ and $\langle R_e^2\rangle$ obey the expected
scaling laws (\ref{Re_Rg_scl}) reasonably well yielding an estimate
for the Flory exponent falling into the interval of $0.58<\nu<0.6$.
This is centered around the best known estimate, $\nu=0.588$,
obtained by means of the renormalisation group approach
\cite{Guida1998}.

As far as the asphericity $A$ and the prolateness $S$ of a polymer
chain are defined \textit{via} combinations of eigenvalues
$\lambda_{\alpha}$ of the gyration tensor $\vvec{Q}$ [see,
Eq.~\ref{ASg_def}], the scaling properties of $\lambda_{\alpha}$ are
also of much interest. Such an analysis for the case of the SAW was
performed first in Ref.~\cite{Sciutto1996} using lattice MC
simulations. It has been demonstrated that the eigenvalues
$\lambda_{\alpha}$ obey the same scaling laws as the traditional
global observables, i.e. $R_e$ and $R_g$
\begin{equation}\label{lambda_scl}
\langle\lambda_\alpha\rangle \sim N^{2\nu},\;\;\; \alpha=1,2,3,
\end{equation}
and that the corrections to scaling for the eigenvalues are
different but have the same sign as their counterparts for $R_g$. In
our DPD simulations we also find that $\lambda_{\alpha}$ obey the
same scaling laws as $R_g$, see, respective legends in
Fig.~\ref{scaling_L123Rgsq} (a). The respective exponents $\nu$ for
each eigenvalue, obtained \textit{via} linear fit of the data for
$N\geq 10$, are indicated in the same figure and are found to be
within the interval $0.574-0.592$. This spread of values must be
attributed to different magnitudes of the correction to scaling
terms \cite{Sciutto1996,Ilnytskyi2007} in each case. These terms are
not accounted for in the current study, as far as accurate values
for $\nu$ are not the primary goal of this study. Therefore, within
the accuracy of our simulations, one may state that the same scaling
properties are obeyed by all eigenvalues $\lambda_{\alpha}$ of the
gyration tensor $\vvec{Q}$. This can be interpreted as an
isotropicity of the self-similarity properties of a polymer chain.

Due to this statement, and given the definitions for $A$ and $S$ (\ref{ASg_def}),
these shape characteristics are expected to be independent of $N$.
To have an additional numerical confirmation for this to hold, we consider the respective nominators and denominators
\begin{eqnarray}\label{AS_nom_den}
A\idx{nom} = \sum_{i=1}^{3}(\lambda_{i}-\bar{\lambda})^{2},
\hspace{10mm} && S\idx{nom} =
\prod^{3}_{i=3}(\lambda_{i}-\bar{\lambda}),\\ A\idx{den} =
6\bar{\lambda}^{2}, \hspace{10mm} && S\idx{den} = \bar{\lambda}^{3},
\end{eqnarray}
denoted here as ``nom'' and ``den''. Taking into account Eq.~(\ref{lambda_scl}), one expects the following scaling
behaviour
\begin{equation}\label{nom_den_scl}
\langle A\idx{nom} \rangle \sim \langle A\idx{den} \rangle \sim N^{4\nu},\;\;\;
\langle S\idx{nom} \rangle \sim \langle S\idx{den} \rangle \sim N^{6\nu}.
\end{equation}
The data shown in Fig.~\ref{scaling_L123Rgsq} (b) indicates that
this scaling behaviour holds well for $N\geq 10$ for all four
properties defined in Eq.~(\ref{AS_nom_den}) yielding the values of
the Flory exponent $\nu$ (see the figure) consistent with their
counterparts for $R_g$ and $\lambda_{\alpha}$ shown in
Fig.~\ref{scaling_L123Rgsq} (a).

\begin{figure}[!h]
\centering
\includegraphics[width=8cm,angle=270]{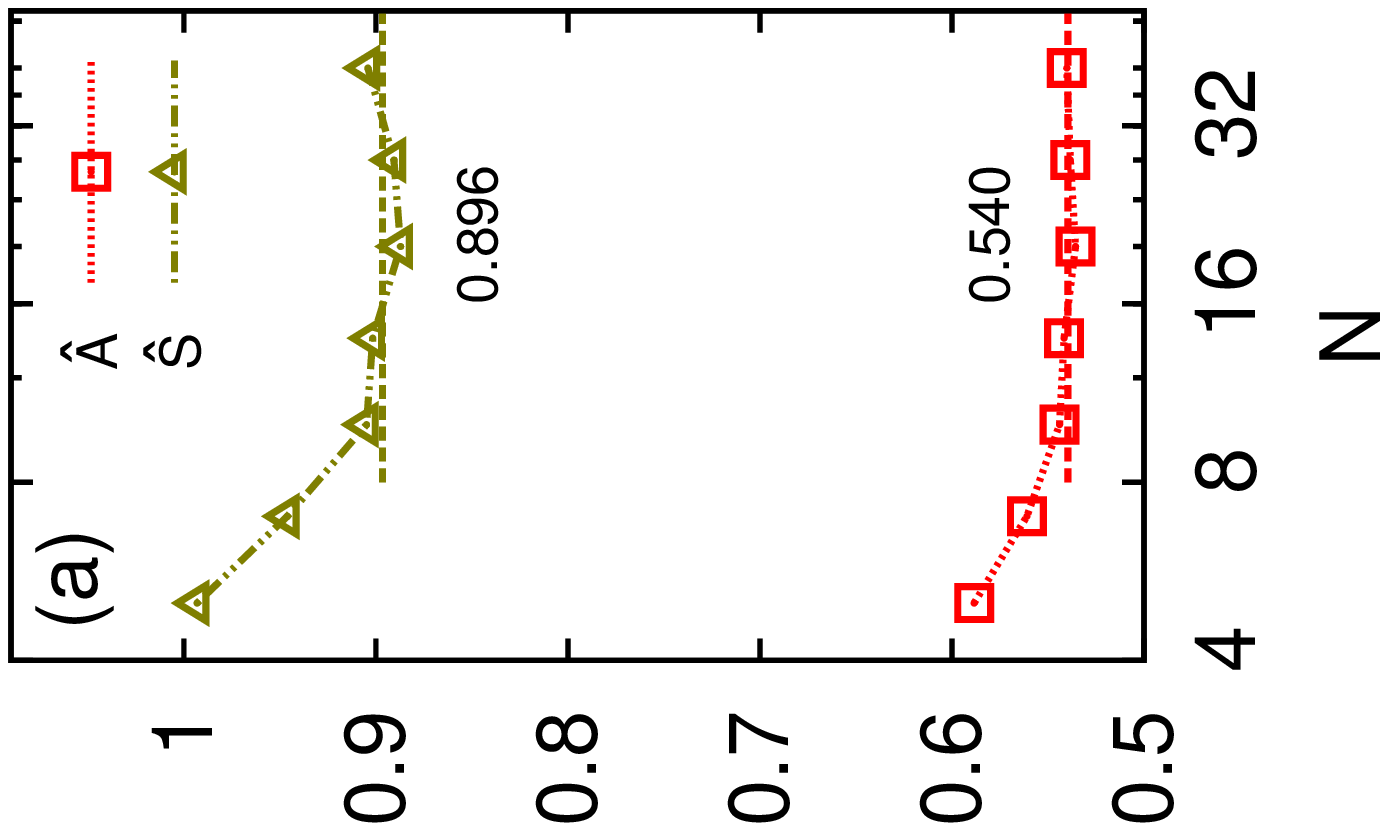}\hspace{-3em}
\includegraphics[width=8cm,angle=270]{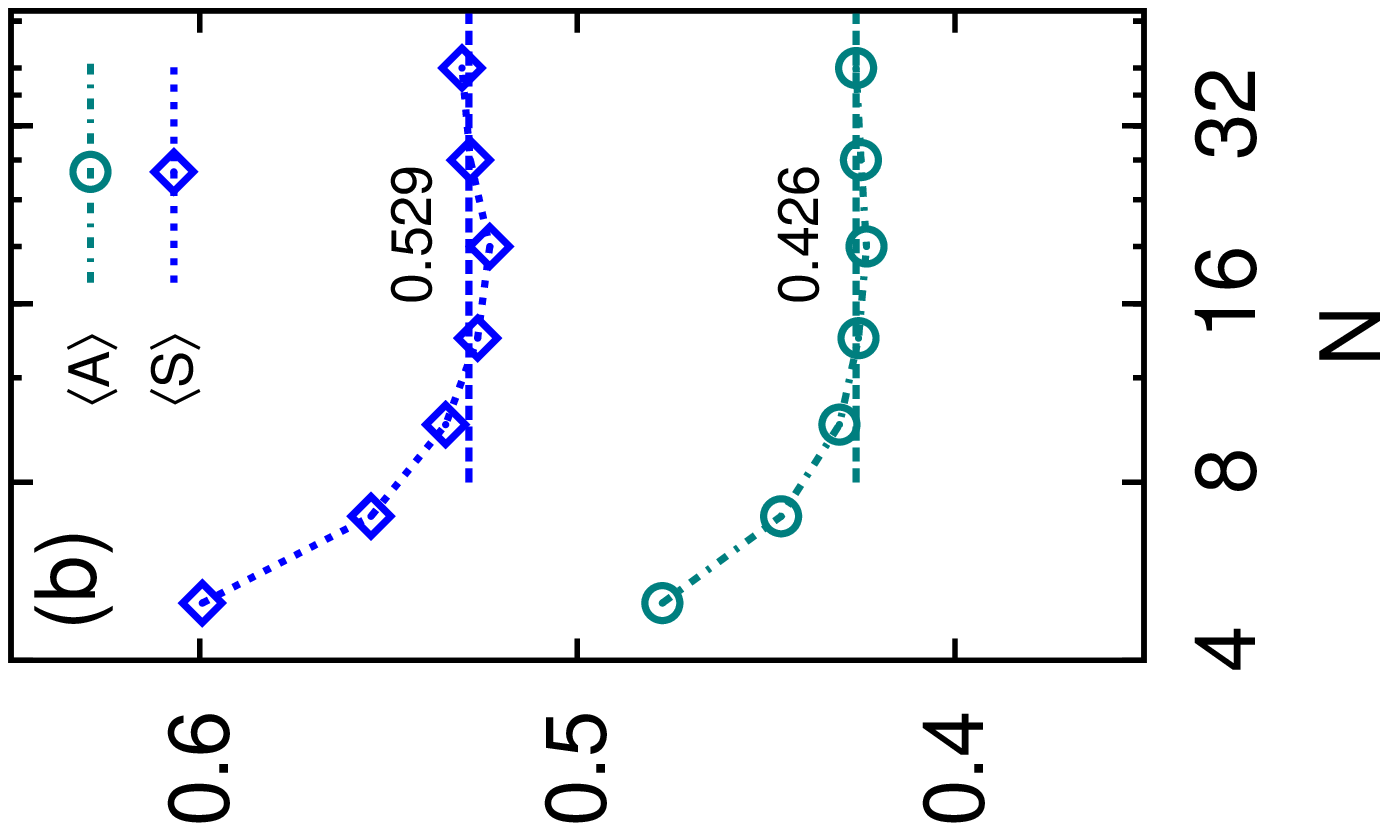}
\caption{\label{scaling_AS}(a) Dependence of the average values
$\hat{A}$ and $\hat{S}$ on $N$. (b) The same for $\langle A\rangle$ and
$\langle S\rangle$. Dashed lines indicate the interval of the averaging
for evaluating the final mean value.}
\end{figure}
As a consequence of this, the average values $\hat{A}$, $\hat{S}$,
$\langle A\rangle$ and $\langle S\rangle$ all demonstrate a weak
dependence on $N$ for $N\geq 10$, as shown in Fig.~\ref{scaling_AS}.
The final mean values for these characteristics are, therefore,
obtained by averaging the data within an interval $10\leq N\leq 40$,
as indicated by dashed lines in both frames of this figure.

\begin{figure}[!h]
\centering
\includegraphics[width=8cm,angle=270]{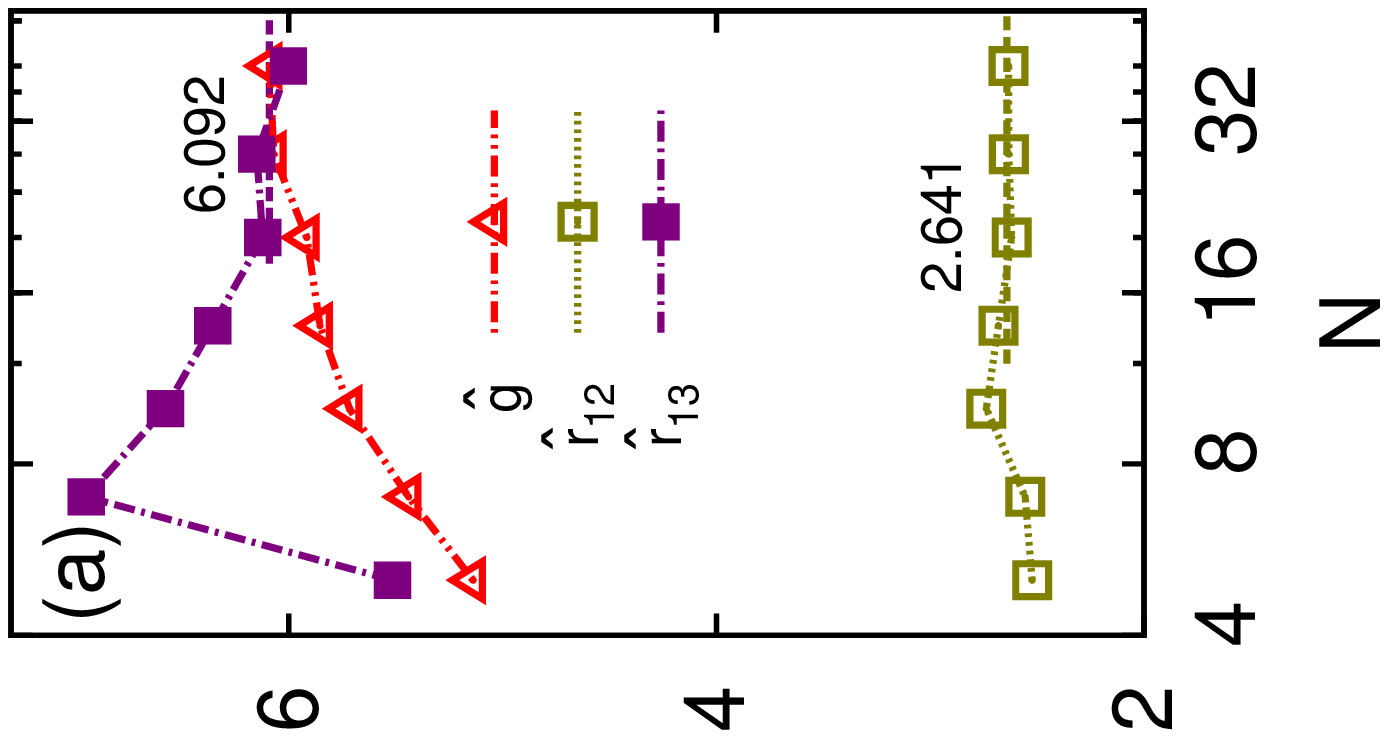}\hspace{-3em}
\includegraphics[width=8cm,angle=270]{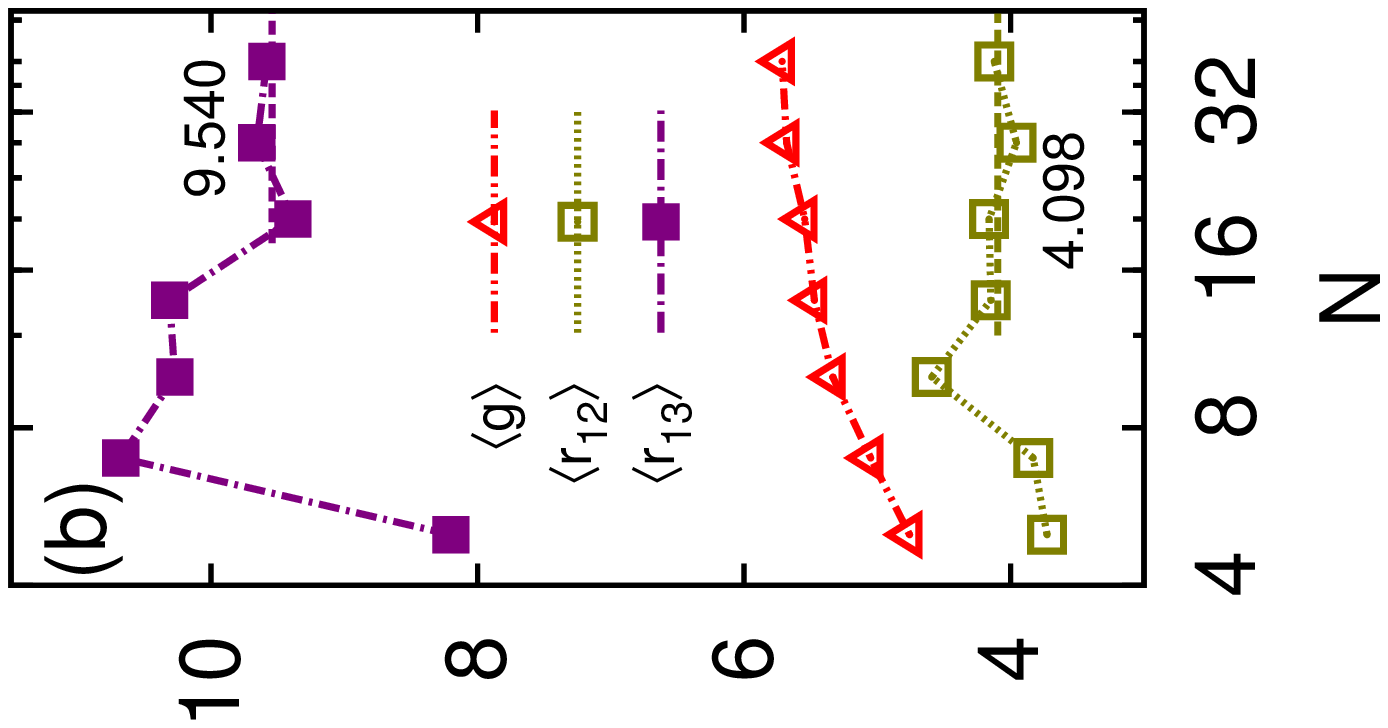}
\caption{\label{scaling_gr12r13}(a) Dependence of the average values
$\hat{g}$, $\hat{r_{12}}$ and $\hat{r_{13}}$ on $N$. (b) The same for $\langle g\rangle$, $\langle r_{12}\rangle$ and $\langle r_{13}\rangle$. Dashed lines indicate the interval of the averaging
for evaluating the final mean value (where applies).}
\end{figure}
The situation is markedly different for other shape characteristics,
where essential $N$ dependence is observed, see
Fig.~\ref{scaling_gr12r13}. The values for $\hat{g}$ and $\langle
g\rangle$ demonstrate steady monotonic growth within the whole
interval $5\leq N \leq 40$ considered in this study. $\hat{r}_{13}$
and $\langle r_{13}\rangle$ stabilize more or less only at $20\leq
N\leq 40$, where the estimate for an average is made. The interval
for $\hat{r}_{12}$ and $\langle r_{12}\rangle$ is a slightly
broader: $14\leq N\leq 40$.

The mean values for the shape properties defined in
Eqs.~(\ref{ASg_hat})-(\ref{r12_r13_av}) and evaluated in our study
as discussed above, are collected in column DPD of
Table.~\ref{tab_aver}. Here, we also list the results obtained for
SAWs by means of other methods, namely, the direct renormalisation
approach [RG (DR)] and Monte Carlo studies (MC), as well as the
respective values for the case of random walks (RW). Let us consider
$\hat{A}$ and $\hat{S}$ first. One should note that the previous
results reported for these characteristics for the case of the SAW,
are found to be very close to their counterparts for the RW case.
Therefore, a high accuracy is needed to distinguish between both
sets unambiguously. In our case of an off-lattice DPD simulations
with explicit solvent, this would be very computationally demanding.
Nevertheless, within the accuracy limitation of this study, the
values found for both $\hat{A}$ and $\hat{S}$ agree reasonably well
with the best respective estimates made for the case of SAWs. For
the case of $\langle A\rangle$ and $\langle S\rangle$, the
difference between the respective values for SAW and RW are more
essential. We find reasonably good agreement between our results and
other data for SAW in as demonstrated in Fig.~\ref{scaling_gr12r13}.

\begin{table}
\begin{centering}
\begin{tabular}{|c|l|l|l|l|l|}
\hline
                        & \multicolumn{4}{|c|}{SAW} & RW \\
\hline
property                &   DPD        & RG    &  MC   & MC$^f$    & \\
                        &              & (DR)  &       &           & \\
\hline
$\hat{A}$               & 0.540        & 0.529$^a$ & 0.546$^c$ & 0.54725 & 0.526$^c$ \\
$\hat{S}$               & 0.896        & 0.893$^a$ &       & 0.91331 & 0.887$^a$ \\
$\hat{g}$               & $>$ 6        & 6.258$^d$ & 6.249$^b$ &         & 6$^a$ \\
$\hat{r}_{12}$          & 2.641        &       &       &         & 2.546 \\
$\hat{r}_{13}$          & 6.092        &       &       &         & 5.657 \\
\hline
$\langle A\rangle$      & 0.426        & 0.415$^b$ & 0.431$^b$ & 0.43337 & 0.394$^e$ \\
$\langle S\rangle$      & 0.529        &       & 0.541$^b$ & 0.54474 & 0.475$^b$ \\
$\langle g\rangle$      & $\leq$ 6     &       &       & & \\
$\langle r_{12}\rangle$ & 4.098        &       &       & & \\
$\langle r_{13}\rangle$ & 9.540        &       &       & & \\
\hline
$\bar{A}$               & 0.418 [G]     &\multicolumn{4}{|c|}{}\\
                        & 0.390 [L]     &\multicolumn{4}{|c|}{}\\
                        & 0.386 [L$'$]  &\multicolumn{4}{|c|}{}\\
$\bar{g}$               & 5.333 [G]     &\multicolumn{4}{|c|}{}\\
                        & 5.444 [L]     &\multicolumn{4}{|c|}{}\\
$\bar{r}_{12}$          & 1.578 [L]     &\multicolumn{4}{|c|}{}\\
$\bar{r}_{13}$          & 3.763 [L]     &\multicolumn{4}{|c|}{}\\
\hline
\end{tabular}
\caption{\label{tab_aver}The results for mean values of asphericity
$\hat{A}$, $\langle A\rangle$, prolateness $\hat{S}$, $\langle
S\rangle$ and size ratios $\hat{g}$, $\langle g\rangle$,
$\hat{r_{12}}$, $\langle r_{12}\rangle$, $\hat{r_{13}}$, $\langle
r_{13}\rangle$ defined according to
Eqs.~(\ref{ASg_hat})-(\ref{r12_r13_av}) and their respective most
probable values $\bar{A}$, $\bar{g}$, $\bar{r}_{12}$ and
$\bar{r}_{13}$. The abbreviations stand for: random walk (RW), self
avoiding walk (SAW), lattice Monte Carlo (MC), direct
renormalization (DR), dissipative particle dynamics (DPD) (this
work). Shorthands for citations: $^a$ \cite{Aronovitz1986}, $^b$
\cite{Jagodzinski1992}, $^c$ \cite{Bishop1988}, $^d$
\cite{Benhamou1985}, $^e$ \cite{Diehl1989}, $^f$
\cite{Zifferer1999a}. Data obtained \textit{via} fits by
Eqs.~(\ref{Gauss}), (\ref{Lhuillier}) and (\ref{Lhuillier2}) are
denoted as [G], [L] and [L$'$], respectively.}
\end{centering}
\end{table}
%

\section{Shape characteristics probability distributions}\label{IV}

It has been observed before
\cite{Bishop1988,Jagodzinski1992,Sciutto1994,Sciutto1996,Zifferer1999a,Zifferer1999b} that the probability distributions of the shape characteristics of a polymer chain
are broad and skewed. This is true for both cases of the RW and SAW and `` implies that any description of the shapes
of random  walks  which is based  only on  mean values of related  magnitudes is incomplete'' \cite{Sciutto1994}.
One of the ways to extend the analysis of the shape characteristics is
to complement the mean values by additional characteristic
values obtained from the respective probability distributions.

\begin{figure}[!h]
\centering
\includegraphics[width=5.5cm,angle=270]{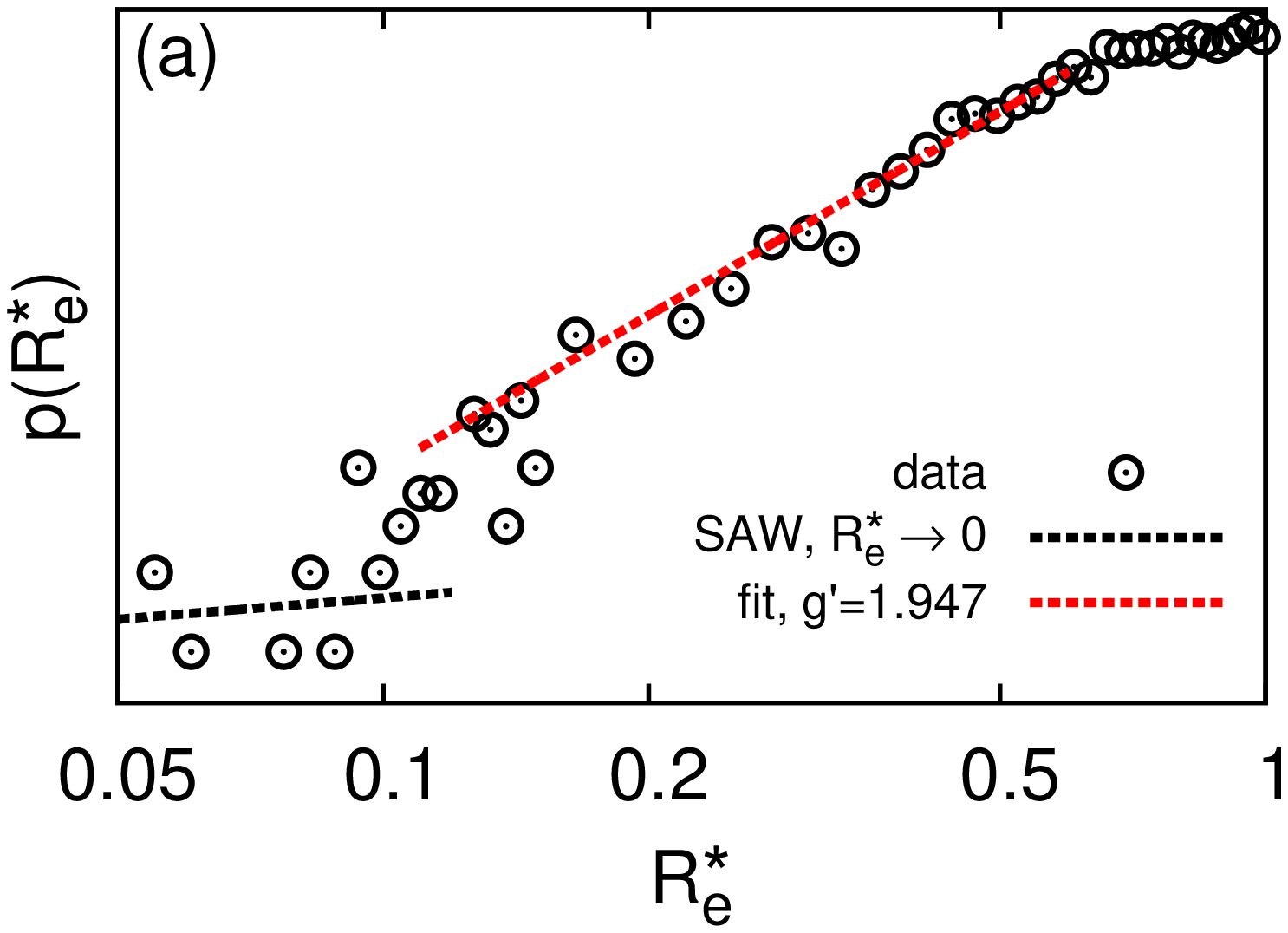}\hspace{-1em}
\includegraphics[width=5.5cm,angle=270]{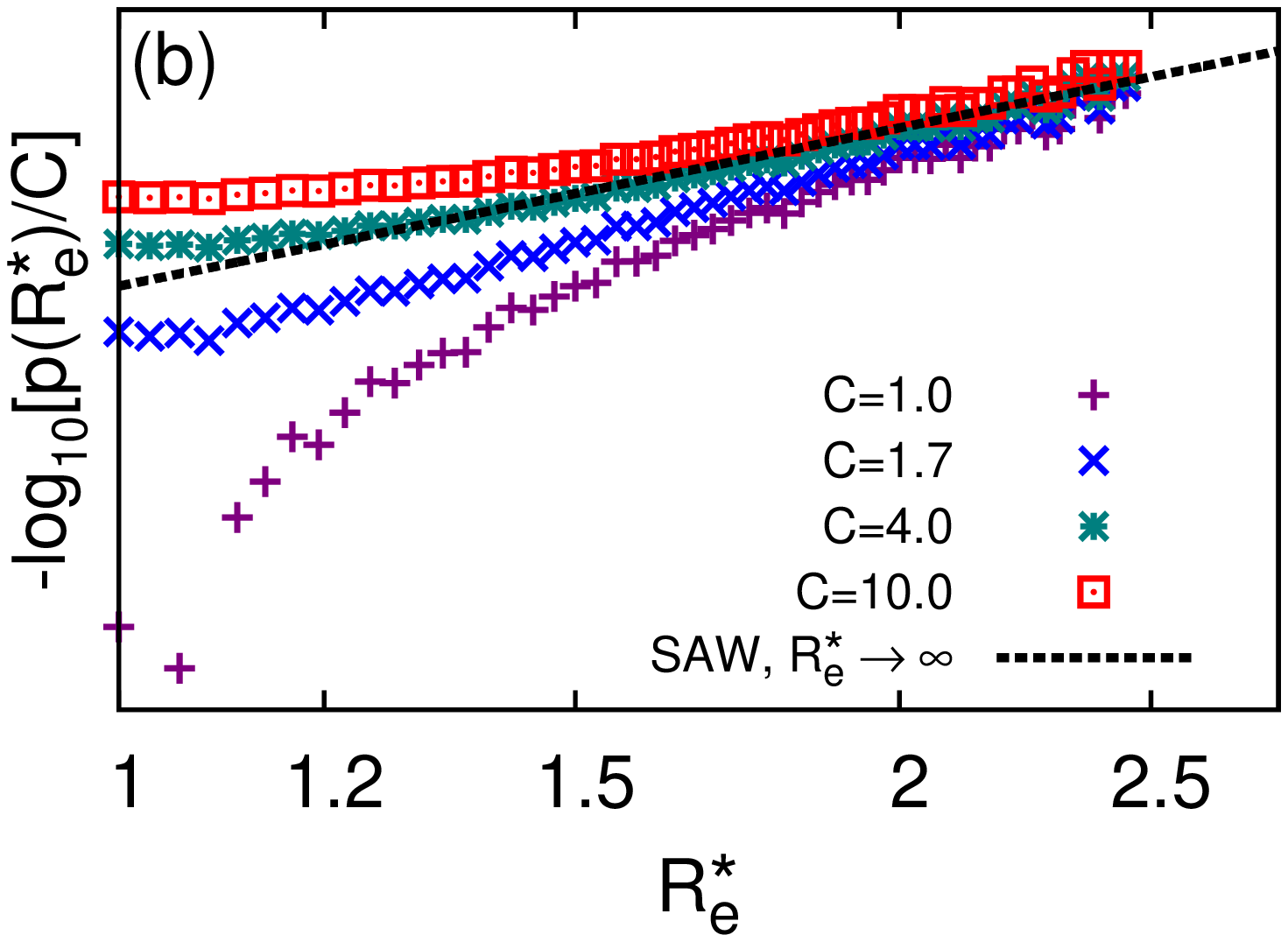}
\caption{\label{Re_dist_LR}Probability distribution $p(R_e^*)$ for
reduced end-to-end distance $R_e^*$. (a) log-log plot for fitting
the asymptotics at $R_e^*\to 0$. (b) Double log plot for fitting the
asymptotics at $R_e^*\gg 0$ built for various trial constant $C$,
see Eq.~(\ref{Re_dist_asympt}) and the text for more details.}
\end{figure}
Before we proceed to the probability distributions for the shape
characteristics, let us check known asymptotics for the
distributions of the end-to-end distance and the gyration radius. To
this end we introduce respective reduced properties
\begin{equation}\label{ReRg_star}
R_e^* =\frac{R_e}{\langle R_e\rangle},\;\;\;
R_g^* =\frac{R_g}{\langle R_g\rangle}.
\end{equation}
As has been shown in \cite{deGennes1979,desCloizeaux1982}, the
probability distribution $p(R_e^*)$ has the following asymptotics:
\begin{equation}\label{Re_dist_asympt}
 p(R_e^*)\approx\left\{
\begin{tabular}{ll}
$\displaystyle A (R_e^*)^{\frac{\gamma-1}{\nu}}$, &~~at~~$R_e^*\to 0$,\\
$\displaystyle C \exp\left [-(R_e^*)^{\frac{1}{1-\nu}}\right]$, &~~at~~$R_e^*\gg 0$.
\end{tabular}
\right.
\end{equation}
where $A$ and $C$ are (non-universal) constants, and $\gamma\approx
1.16$. We built a single cumulative histogram for the probability
distribution $p(R_e^*)$ based on simulation data within a scaling
regime, i.e. $10\leq N\leq 40$. To examine its asymptotics at
$R_e^*\to 0$, it was plotted in a log-log scale, see
Fig.~\ref{Re_dist_LR} (a). The $\frac{\gamma-1}{\nu}$ power law is
displayed \textit{via} dashed line and is marked as $R_e^*\to 0$
label. It holds approximately but suffers from somewhat insufficient
accuracy near the tail of the distribution (where the statistics is
the poorest). However, one also observes a large region of $R_e^*$
values where another power law holds well, namely $(R_e^*)^{g'}$
with $g'\approx 2$. To check for the asymptotics of $p(R_e^*)$ at
$R_e^*\gg 0$, the logarithm of the histogram $p(R_e^*)$ is plotted
now scaled by the factor $C$. As this factor is not known \textit{a
priori}, we plotted a family of curves using various values for $C$,
see Fig.~\ref{Re_dist_LR} (b). All curves converge to the
exponential asymptotics in Eq.~(\ref{Re_dist_asympt}), which is
shown in the figure \textit{via} a dashed black line. One may
conclude therefore that the known asymptotics for the probability
distribution $p(R_e^*)$ are adequately reproduced in our
simulations.

\begin{figure}[!h]
\centering
\includegraphics[width=7cm,angle=270]{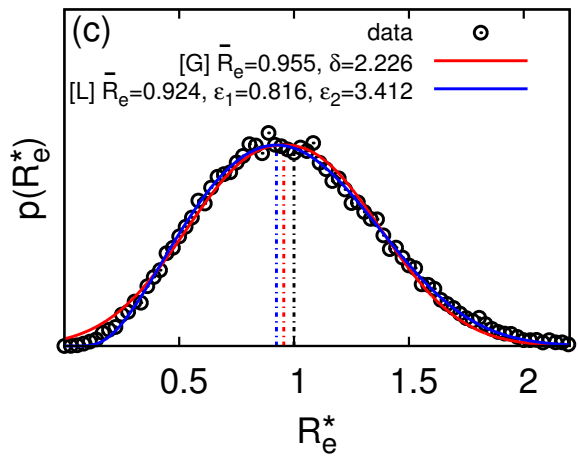}\\
\caption{\label{Re_dist}Probability distribution $p(R_e^*)$
and the results of fits according to Eq.~(\ref{Gauss}) (marked as
[G]) and Eq.~(\ref{Lhuillier}) (marked as [L]). The parameters
of fits are shown in a figure.}
\end{figure}
One should note that the asymptotic regimes (\ref{Re_dist_asympt})
are observed rather at extreme values of $R_e^*$. Therefore, it is
of interest whether or not the probability distribution $p(R_e^*)$
can be fitted \textit{via} appropriate analytic expressions
reasonably well within a whole interval of $R_e^*$ values. The form
for $p(R_e^*)$, as seen in Fig.~\ref{Re_dist}, appears to be weakly
asymmetric. Therefore, one of the obvious analytic expressions to
apply is a generalised Gaussian one
\begin{equation}\label{Gauss}
p_G(x) = A \exp \left[-\left(\frac{x-x_0}{\sigma_0}\right)^\delta\right],\;\;\;\bar{x}=x_0.
\end{equation}
Here and thereafter, the most probable value is denoted as
$\bar{x}$. This fit is marked as [G] in Fig.~\ref{Re_dist} and
yields the exponent $\delta=2.226$ and the most probable value of
$\bar{R}_e=0.955$. One should remark that this fit closely resembles
a standard Gaussian distribution (achieved for parameters values
$\delta=2$ and $\bar{R}_e=1$). Another choice can be terms of a
Lhuillier-like form (for more details, see below)
\begin{equation}\label{Lhuillier}
p_L(x) = B \exp \left[-\Big(\frac{x'}{x}\Big)^{\epsilon_1} - \Big(\frac{x}{x'}\Big)^{\epsilon_2}\right],\;\;\;
\bar{x}=x'\Big(\frac{\epsilon_1}{\epsilon_2}\Big)^{\frac{1}{\epsilon_1+\epsilon_2}}.
\end{equation}
This fit is marked as [L] in Fig.~\ref{Re_dist} yielding
$\epsilon_1=0.816$, $\epsilon_2=3.412$ and $\bar{R}_e=0.924$. One
may conclude that the values for $\bar{R}_e$ obtained in both fits
are very close to the mean value for the scaled end-to-end distance
which is equal to one.

Most studies on probability distributions of shape characteristics
of polymer chains are undertaken for the case of a Gaussian chain,
which mimics the RW
\cite{Solc1971a,Solc1971b,Rudnick1986,Gaspari1987,Wei1990,Sciutto1994,Sciutto1995}.
In this case the analytic evaluation is possible, as well as the
$1/d$ expansion, all leading to certain analytic expressions for the
probability distributions $p(\lambda_{\alpha})$ and, in some cases,
for $p(A)$ \cite{Sciutto1994,Sciutto1995}. On the other hand, in
most studies that address such probability distributions for the SAW
case
\cite{Jagodzinski1992,Zifferer1999a,Zifferer1999b,Blavatska2011},
the authors mainly concentrate on broad shapes of such distributions
and compare these with their counterparts for the RW. While an
analytic solution for the SAW is not available  , there are several
indirect options suggesting certain analytic expressions for the
probability distributions of interest. One path is suggested by
Sciutto \cite{Sciutto1996}, where the chi-squared form of the
distributions $p(\lambda_{\alpha})$ are obtained analytically for
the RW is transferred to the SAW case with the different set of
parameter values. While the result is very satisfactory for the
eigenvalues ratios $\lambda_{\alpha}/\lambda_{\beta}$, the
probability distributions $p(A)$ and $p(S)$ are not fitted
accurately.

\begin{figure}[!h]
\centering
\includegraphics[width=7cm,angle=270]{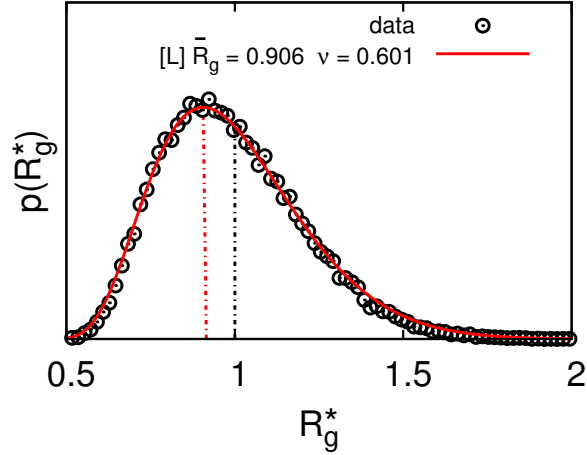}\\
\caption{\label{Rg_dist}Probability distribution $p(R_g^*)$ and the
results of a fit according to Eq.~(\ref{Lhuillier}) with the
exponents $\alpha_1=\frac{3}{3\nu-1}$ and
$\alpha_2=\frac{1}{1-\nu}$. Most probable value and Flory exponent
$\nu$ as the result of a fit are shown.}
\end{figure}
Another approach has been suggested by Lhuillier
\cite{Lhuillier1988}, where the empirical form (\ref{Lhuillier}) has
been deduced for the probability distribution $p(R_g^*)$. In
particular, the authors consider statistical weights for extreme
cases of collapsed ($R_g^* \to 0$) and highly stretched ($R_g^* \gg
1$) conformations resulting in the form (\ref{Lhuillier}) for the
probability distribution $p(R_g^*)$. Both exponents
$\alpha_1=\frac{3}{3\nu-1}$ and $\alpha_2=\frac{1}{1-\nu}$ are found
to depend solely on the Flory exponent $\nu$ (for the
three-dimensional case). This, therefore, opens up a possibility to
have another independent estimate for the exponent $\nu$ performing
a fit of $p(R_g^*)$ obtained in simulations to the form
(\ref{Lhuillier}) taking into account the expressions for $\alpha_1$
and $\alpha_2$ as functions of $\nu$. Applying this analytic
expression (\ref{Lhuillier}) to our simulation data we find it to
work extremely well, see Fig.~\ref{Rg_dist}, yielding the estimate
$\nu=0.601$ for the Flory exponent. It is consistent with previous
estimates provided in Figs.\ref{scaling_L123Rgsq} and
\ref{scaling_AS}. The most probable value is $\bar{R}_g=0.906$ which
deviates for about 10\% from the mean value equal to one, as the
result of an essential asymmetry of the distribution. Due to this
asymmetry, the fit used in Eq.(\ref{Gauss}) makes essentially no
sense.

\begin{figure}[!h]
\centering
\includegraphics[width=5cm,angle=270]{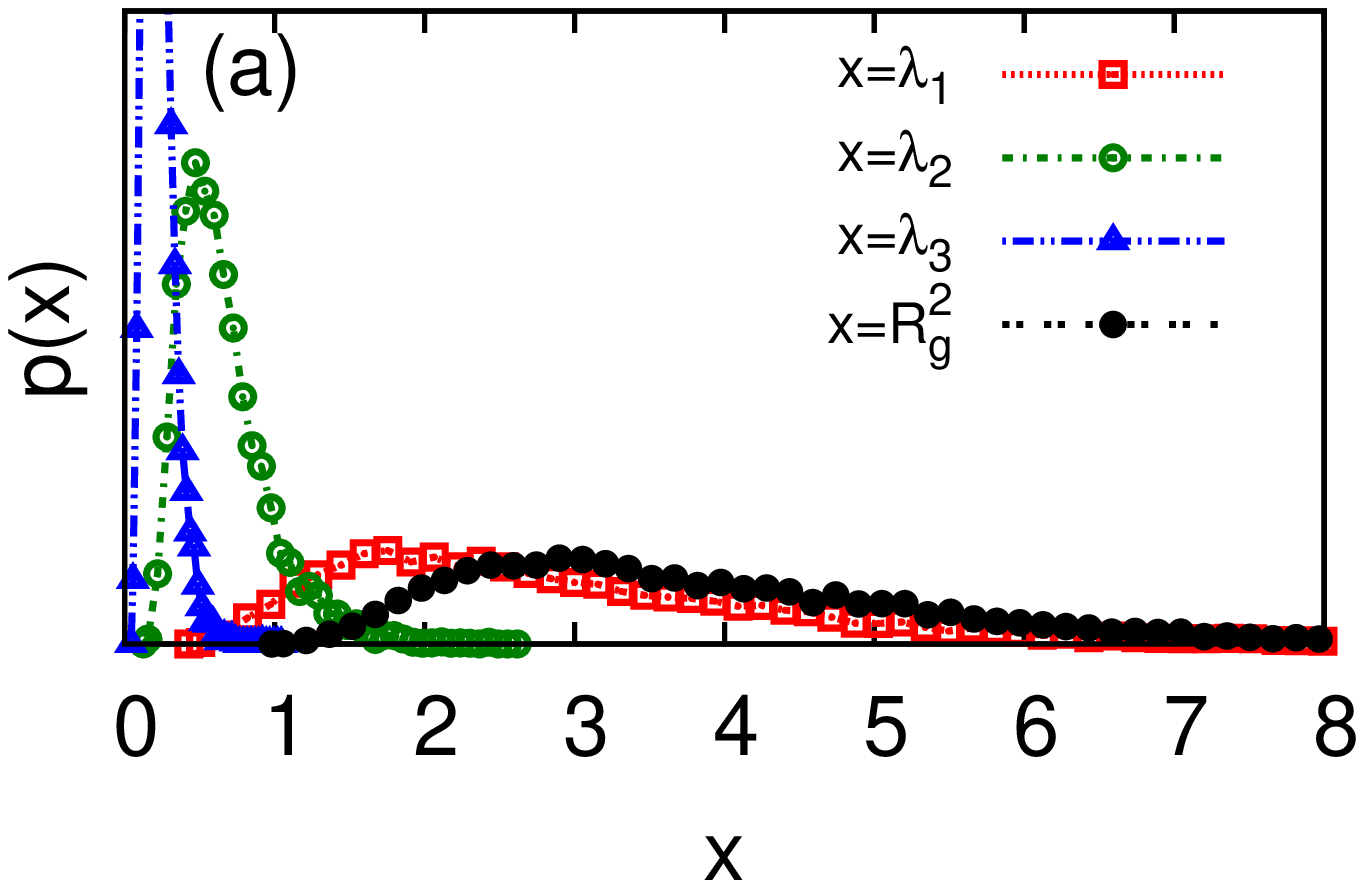}\hspace{-1em}
\includegraphics[width=5cm,angle=270]{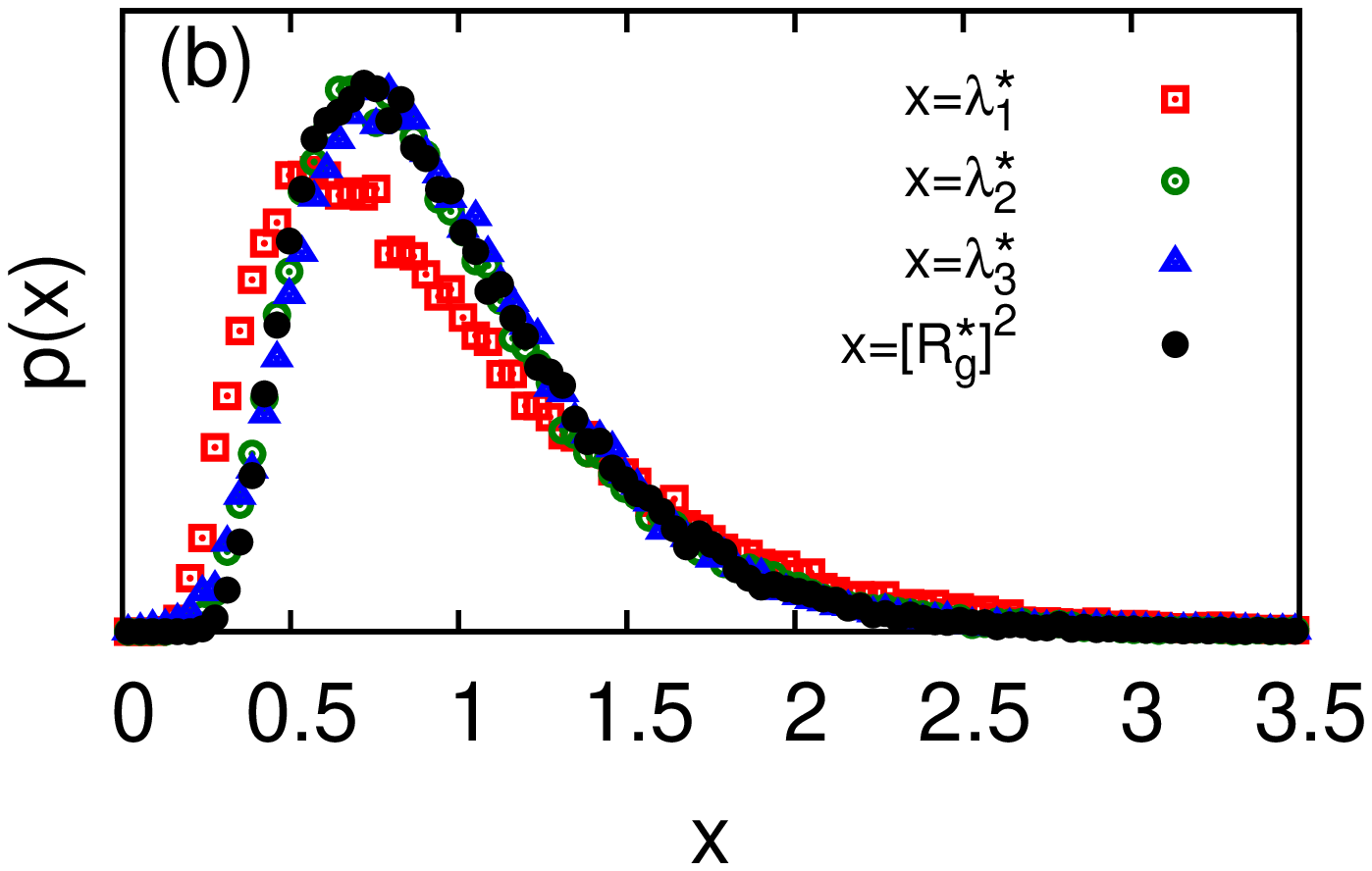}
\caption{\label{L123_dist}Probability distributions for the eigenvalues
$\lambda_{\alpha}$ and squared radius of gyration $R_g$. (a) Distributions
for unscaled variables. (b) Distributions for scaled variables, $x^*=x/\langle x\rangle$.}
\end{figure}
In a further approach, one may consider the probability
distributions for the eigenvalues $\lambda_\alpha$ of the gyration
tensor $\vvec{Q}$ shown in Fig.~\ref{L123_dist} (a). For the sake of
comparison, we also display the probability distribution for their
sum, which is equal to $R_g^2$. One should note an increase
width of the distributions corresponding to an increase of
respective eigenvalues $\lambda_1>\lambda_2>\lambda_3$. This fact
has been already discussed in a number of studies
\cite{Jagodzinski1992,Zifferer1999a,Zifferer1999b,Blavatska2011}.
Another important point whether or not the probability distributions
for their scaled counterparts
$\lambda^*_{\alpha}=\lambda_{\alpha}/\langle\lambda_{\alpha}\rangle,
\alpha=1,2,3$ can be mapped onto a single master curve has not
discussed so far, except for the case of the RW \cite{Solc1971a}.
There the distributions $p(\lambda^*_{\alpha})$ are found not to
coincide. Our respective simulation data are displayed in
Fig.~\ref{L123_dist} (b), from which it appears that: (i) the
probability distributions $p(\lambda^*_2)$, $p(\lambda^*_3)$ and
$p[(R_g^*)^2]$ overlap very closely with little or none
deviation, and (ii) while  $p(\lambda^*_1)$ does not. Currently, we
do not have single physical explanation for point (i) but point (ii)
can be explained easily. Let us go back to Fig.~\ref{L123_dist} (a).
In a limit of a highly stretched chain one has
$\lambda_2,\lambda_3\to 0$ and $R_g^2\approx \lambda_1\geq 1$.
Hence, at large enough $R_g^2$ and $\lambda_1$, their respective
distributions should overlap. On the other hand, in a most probable,
coil state, the contribution to $R_g^2$ from $\lambda_2$ and
$\lambda_3$ is essentially non-zero. Hence, $R_g^2>\lambda_1$ and
the maximum position for $p(R_g^2)$ is shifted to larger values as
compared to that for $p(\lambda_1)$, as observed in
Fig.~\ref{L123_dist} (a). Due to these two requirements, the
distributions $p(R_g^2)$ and $p(\lambda_1)$ can not be reduced to
the same  master curve by the scaling transformation only. We find,
however, that the $p(\lambda^*_1)$ distribution can be matched with
the others by applying a shift $\lambda_1 \to \lambda_1-\langle
\lambda_2+\lambda_3\rangle$, which is not surprising, as it turns
the shifted value into some approximation of the $R_g$ (this is not
shown in the figure for a sake of brevity). Nevertheless, one may
report relative proximity of the probability distributions for the
scaled properties $\lambda^*_{\alpha},\;\alpha=1,2,3$ and
$(R_g^*)^2$ indicating strong isotropicity of the polymer chain
along with its eigenvalues. Indeed, not only the mean values for all
eigenvalues scale by the same law [Fig.~\ref{scaling_L123Rgsq} (a)],
but also their probability distributions do so to great extent.

\begin{figure}[!h]
\centering
\includegraphics[width=7cm,angle=270]{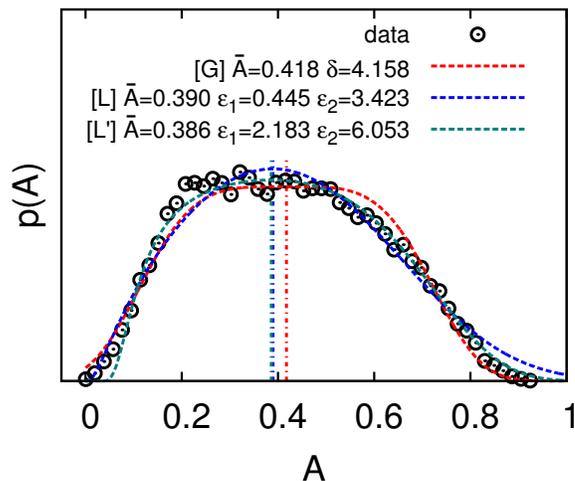}
\caption{\label{A_dist}Probability distribution $p(A)$ and the
results of fits according to Eq.~(\ref{Gauss}) (marked as [G]),
Eq.~(\ref{Lhuillier}) (marked as [L]) and Eq.~(\ref{Lhuillier2})
(marked as [L$'$]). The parameters of fits are shown in a figure.}
\end{figure}
The prediction of the asymptotics for the probability distributions
of the shape characteristics similar to the way it is done by
Lhuillier \cite{Lhuillier1988} for $R_g$ faces severe difficulties.
Indeed, the interval of small $R_g$ is associated with a collapsed
chain, whereas that of large $R_g$ -- to highly stretched
conformations. This does not hold for the asphericity $A$, where
small (large) values of $A$ contain contributions from the whole
range of conformations being spherically symmetric (asymmetric). As
was shown in Ref.~\cite{Sciutto1996}, the form of the probability
distributions $p(A)$ and $p(S)$ differ for the cases of SAW and RW.
Therefore, reparametrisation of the chi-squared expressions valid
for a RW do not reproduce well the respective distributions for the
SAW case. In this paper we chose another route: reparametrisation of
the Lhuillier-like expression (\ref{Lhuillier}), heuristically
suggested for the gyration radius distribution, and its extension
for the distribution of shape characteristics.

Similarly to the cases of the $R_e$ and $R_g$ probability
distributions, we build a cumulative probability distribution
histogram for each shape characteristic based on the simulation data
for $N=10$, $14$, $20$, $28$ and $40$. As was shown in
Figs.~\ref{scaling_L123Rgsq}-\ref{scaling_gr12r13}, within this
interval of $N$ the scaling laws follow sufficiently well and the
probability distributions are found to overlap. The study of such a
cumulative histogram essentially enhances the statistics of the
analysed data.

The form of the probability distribution $p(A)$ for the asphericity,
shown in Fig.~\ref{A_dist}, indicates a relatively low asymmetry,
therefore we attempt fits using both functional forms (\ref{Gauss})
and (\ref{Lhuillier}). This yield the following exponents:
$\delta=4.158$, $\epsilon_1=0.445$ and $\epsilon_2=3.423$ and the
most probable values $\bar{A}=0.418$ and $0.390$ for $p_G(A)$ and
$p_L(A)$, respectively. The fit \textit{via} Lhuillier-like form
follows the shape of $p(A)$ more closely. An even better
approximation can be achieved by using the extended Lhuillier-like
form
\begin{equation}\label{Lhuillier2}
p_L'(x) = C \exp \left[-\Big(\frac{x_1}{x}\Big)^{\epsilon_1} - \Big(\frac{x}{x_2}\Big)^{\epsilon_2}\right],\;\;\;
\bar{x}=\Big(x_1^{\epsilon_1}\, x_2^{\epsilon_2}\,\frac{\epsilon_1}{\epsilon_2}\Big)^{\frac{1}{\epsilon_1+\epsilon_2}}.
\end{equation}
In this case the exponents are: $\epsilon_1=2.183$ and
$\epsilon_2=6.053$ and the most probable value for the asphericity
is $\bar{A}=0.386$. This is very close to the one obtained with the
(\ref{Lhuillier}) form. The most probable values $\bar{A}$ are added
to Tab.~\ref{tab_aver} and one notes that the value obtained from
the fit (\ref{Gauss}) is close to the average value $\langle
A\rangle$, whereas both  values obtained \textit{via} fits
(\ref{Lhuillier}) and (\ref{Lhuillier2}) are closer to the
asphericity of a RW.

\begin{figure}[!h]
\centering
\includegraphics[width=6cm,angle=270]{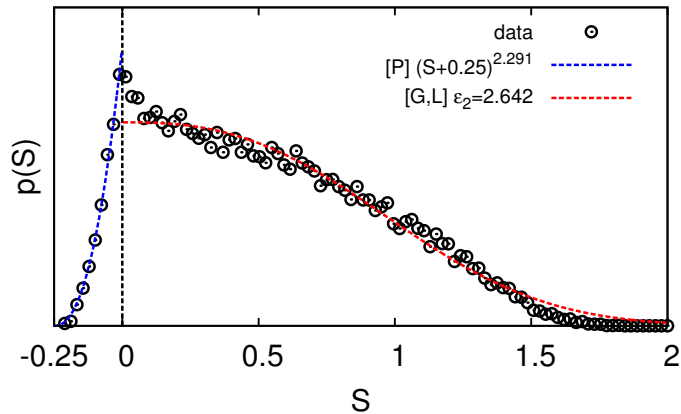}
\caption{\label{S_dist}Probability distribution $p(S)$
and the results of a fit according to Eq.~(\ref{Sfit}). The parameters of a fit are shown in a figure.}
\end{figure}
The prolateness $S$, as already mentioned above, distinguish es
between the oblate ($S<0$) and prolate ($S>0$) conformations of a
chain. The shape of its probability distribution, $p(S)$, is also
markedly different in these two intervals of $S$  values, see
Fig.~\ref{S_dist}. We were unable to fit it by a single analytic
form but instead opted for two separate fits, at $S<0$ and $S>0$
\begin{equation}\label{Sfit}
p(S) =\left\{
\begin{tabular}{ll}
$\displaystyle A(S+0.25)^\beta$, & $S<0$,\\
$\displaystyle B\exp\left[ - \left(\frac{x}{x_0}\right)^{\epsilon_2}\right]$, & $S>0$.
\end{tabular}
\right.
\end{equation}
The analytic form used for $S>0$ can be attributed to a limit case
of both Eq.~(\ref{Gauss}) (at $x_0=0$) and of Eq.~(\ref{Lhuillier2})
(at $x_1=0$), therefore this fit is marked as [G, L$'$ or L] in
Fig.~\ref{S_dist}. One can not define the most probable value
$\bar{S}$ for the prolateness from a fit (\ref{Sfit}).

\begin{figure}[!h]
\centering
\includegraphics[width=7cm,angle=270]{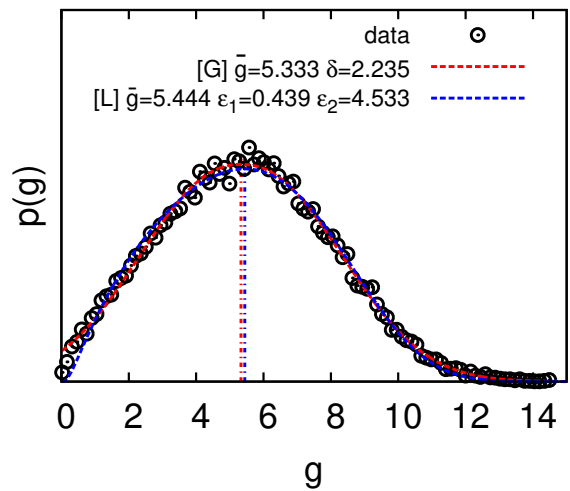}
\caption{\label{g_dist}Probability distribution $p(g)$ and the
results of fits according to Eq.~(\ref{Gauss}) (marked as [G]) and
Eq.~(\ref{Lhuillier}) (marked as [L]). The parameters of fits are
shown in the figure.}
\end{figure}
The probability distribution $p(g)$ for the size ratio $g$ is shown
in Fig.~\ref{g_dist}. It is fitted similarly to the $p(A)$
distribution, \textit{via} analytic forms (\ref{Gauss}) and
(\ref{Lhuillier}). This yields the respective exponents
$\delta=2.235$, $\epsilon_1=0.439$ and $\epsilon_2=4.533$. The most
probable values are $\bar{g}=5.333$ and $5.444$, respectively. One
should note that other approaches evaluate the $\hat{g}$ value only
which is equal to $6$ for the RW and is found to be larger that $6$
for the SAW (see, Tab.~\ref{tab_aver}). It is evident from
Fig.~\ref{scaling_gr12r13} that reliable estimates for both
$\hat{g}$ and $\langle g\rangle$ require the use of a polymer chain
longer than $N=40$ (maximal length used in this study) and,
therefore, can not be provided within this study. Therefore, the
most probable values $\bar{g}$ obtained here do not have any
counterparts to be compared with and can serve as an indication of
high asymmetry of the $p(g)$ distribution only.

\begin{figure}[!h]
\centering
\includegraphics[width=6cm,angle=270]{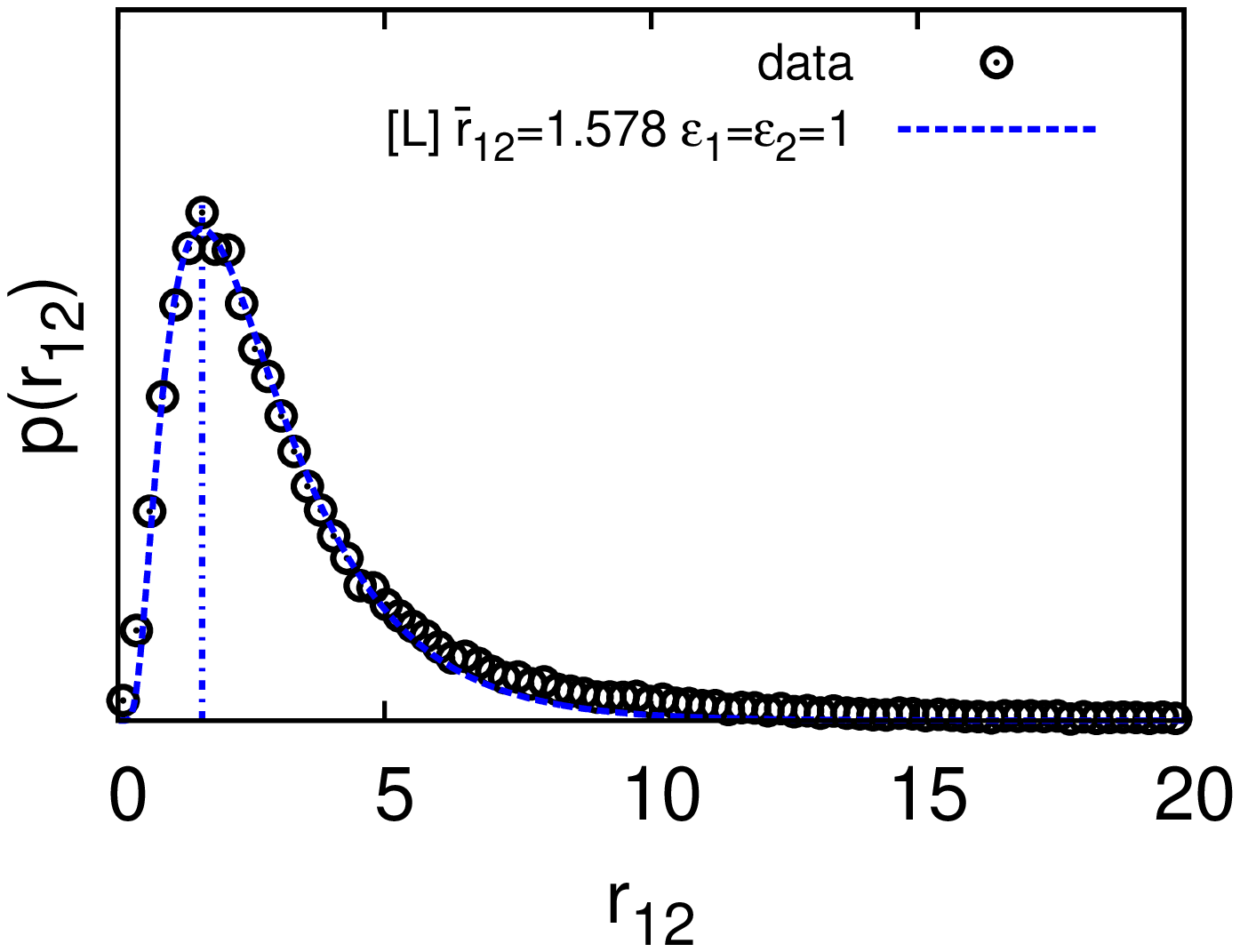}\hspace{-1em}
\includegraphics[width=6cm,angle=270]{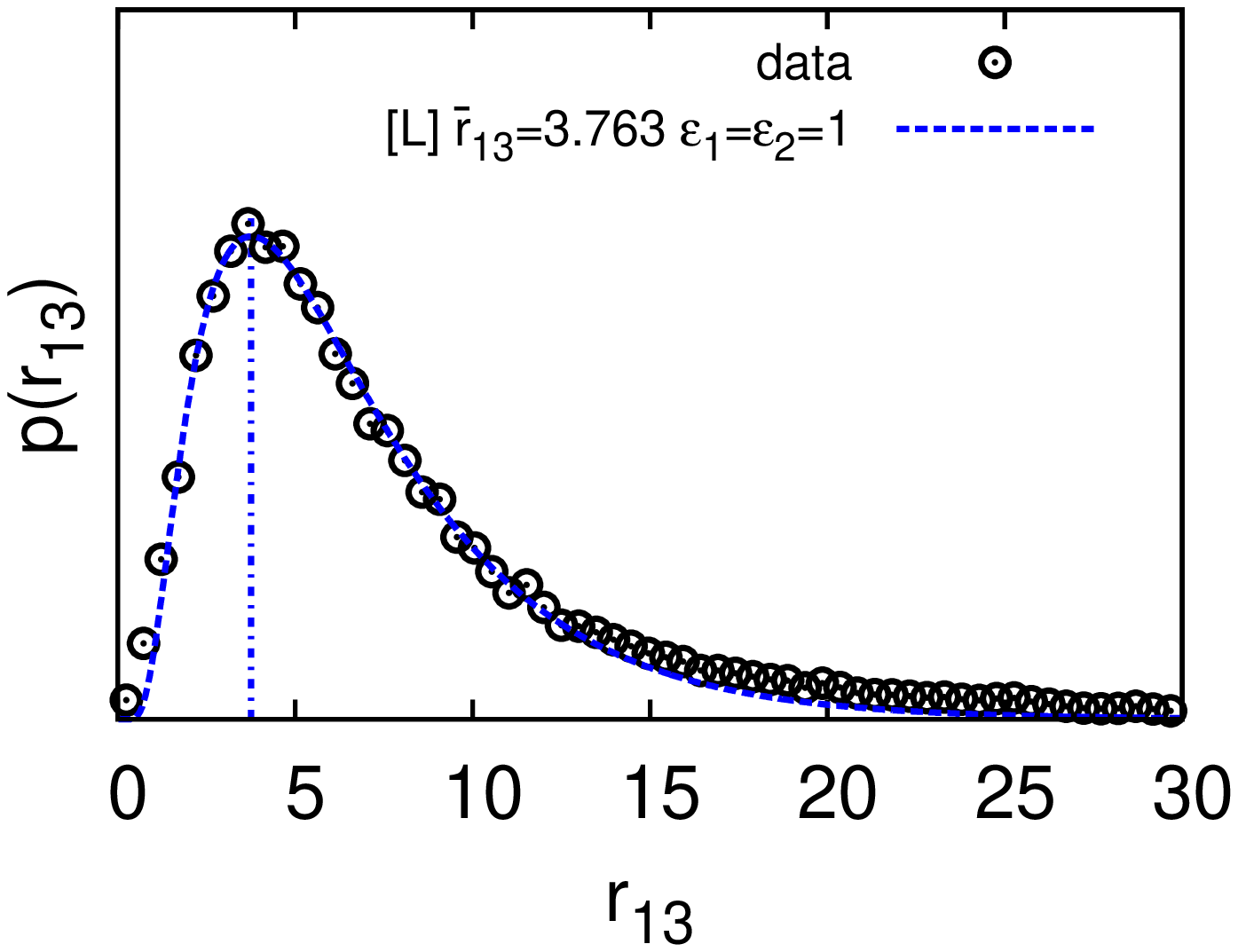}
\caption{\label{r_dist}(a) Probability distribution $p(r_{12})$ and
the results of a fit according to Eq.~(\ref{Lhuillier}) with fixed
$\epsilon_1=\epsilon_2=1$ (marked as [L]). (b) The same for
$p(r_{13})$. The parameters of fits are shown in the figure.}
\end{figure}
The probability distributions $p(r_{12})$ and $p(r_{13})$ appear to
be of a similar, highly asymmetric form, see Fig.~\ref{r_dist} (a)
and (b). Both were preliminarily fitted by the general expression
(\ref{Lhuillier}) yielding the exponents $\epsilon_1$ and
$\epsilon_2$ close to $1$. Therefore, in the final fits we set
$\epsilon_1=\epsilon_2=1$ and obtain the following estimates for the
most probable values: $\bar{r}_{12}=1.578$ and $\bar{r}_{13}=3.763$,
the fits are shown in Fig.~\ref{r_dist}. These values differ about
$2.5-3$ times from their counterparts $\langle r_{12}\rangle$ and
$\langle r_{13}\rangle$ (see, Tab.~\ref{tab_aver}) due to high
asymmetry of respective distributions.

\section{Conclusions}\label{V}

Here we have analysed the shape characteristics of a coarse-grained
polymer chain in a good solvent using DPD simulations. One of the
aims was to check the universality of the properties not directly
connected to the scaling power laws. In this respect we here
evaluated the shape characteristics which can be seen as
counterparts of the universal critical amplitude ratios in critical
phenomena.

As far as the simulations employ an off-lattice model for the
polymer chain and an explicit solvent, these are more
computationally intensive than their lattice Monte Carlo
counterparts and can not match the latter in numerical accuracy.
However, using this simulation technique, one bridges the gap to a
large number of studies targeted on at self-assembly of various
amphi- and polyphilic systems and to the flow-driven phenomena on
the nanoscale. Hence, according to our aim, we rather focus on a
qualitative analysis of the mean and most probable values for the
shape characteristics, as well as on the form of their probability
distributions. Therefore, no rigorous analysis of the numerical
errors was performed and these are not displayed in the figures or
in the table of results.

The origin our analysis is a study of the scaling properties of the
gyration tensor components. We find that, for chain lengths $N\geq
10$, not only the end-to-end distance and gyration radius, but also
all three eigenvalues of the gyration tensor scale with the same
power law. The Flory exponent $\nu$ obtained for each of these
properties independently, within the accuracy of the simulations, is
close to the best theoretical estimate $\nu=0.588$. Based on the
analytic expressions, the shape characteristics of the chain are
expected to be independent of $N$ in the same interval of chain
lengths. This has been checked by plotting each of them against $N$.

The mean values for the shape characteristics were evaluated by
averaging their respective values obtained for chain lengths of
$N=10$, $14$, $20$, $28$ and $40$. The mean values for the
asphericity and prolateness are found to compare well with the
respective results obtained by means of other approaches, whereas
the shape anisotropy requires simulations of longer chains. The
averaging procedure for these properties, however, is found to be
rather ambiguous, as the probability distributions for all shape
characteristics are wide and many are highly asymmetric. Therefore,
alongside with the mean values, one may consider looking at the most
probable values of each characteristic as well. This can be
performed by fitting the probability distribution for each
characteristic by certain analytic expressions and determining the
most probable value based on the positions of its maxima.

Here, we started from the analysis of the probability distributions
for the reduced end-to-end distance and the gyration radius. For the
former case, the known analytic asymptotics of Cloiseaux-de Gennes
were reproduced, whereas for the latter the Lhuillier analytic
expression is found to be very accurate yielding another independent
estimate for the Flory exponent $\nu$. The analytic expressions in
both cases are based on an analysis of the conformation statistics,
which is more difficult to apply to the shape characteristics. For
the latter, we suggest heuristic analytic expressions based on the
Lhuillier form and, for the symmetric distributions, on a
generalised Gaussian distribution. This leads to simple analytical
expressions with the coefficients and exponents found from fitting
the simulation data. This allows us to complement the mean values
for the shape properties by their most probable values found as the
maximal positions of the respective distributions.

The study can be interpreted as yet another validation on the
appropriateness of the use of soft coarse-grained potentials to
describe the self-avoiding polymer chain. Not only the scaling
properties of the end-to-end distance and gyration radius of such
chains are in good agreement with the theoretical data, but also the
mean values of the shape characteristics and their probability
distributions are found to closely mimic the results of more
accurate lattice Monte Carlo simulations.

\section*{Acknowledgements}

We are thankful to Victoria Blavatska for useful discussions. This work was supported in part by FP7 EU IRSES
projects  No. $295302$ ``Statistical Physics in Diverse Realizations", No. $612707$ ``Dynamics of and in Complex
Systems", No. $612669$ ``Structure and Evolution of Complex Systems with Applications in Physics and Life Sciences",
and by the Doctoral College for the Statistical Physics of Complex Systems, Leipzig-Lorraine-Lviv-Coventry
$({\mathbb L}^4)$.
\clearpage

\providecommand{\newblock}{}


\begin{thebibliography}{10}
\expandafter\ifx\csname url\endcsname\relax
  \def\url#1{{\tt #1}}\fi
\expandafter\ifx\csname urlprefix\endcsname\relax\def\urlprefix{URL
}\fi \providecommand{\eprint}[2][]{\url{#2}}

\bibitem{deGennes1979}
de~Gennes P~G 1979 {\em Scaling Concepts in Polymer Physics\/}
(Cornell
  University Press Ithaca London)
  \urlprefix\url{http://dx.doi.org/10.1063/1.2914118}

\bibitem{desCloizeaux1982}
des Cloizeaux J 1982 Theory of polymers in solution {\em Phase
Transitions
  Carg{\`{e}}se 1980\/} (Springer Science + Business Media) pp 371--394
  \urlprefix\url{http://dx.doi.org/10.1007/978-1-4613-3347-0\_16}

\bibitem{Lhuillier1988}
Lhuillier D 1988 {\em J. Phys. France\/} {\bf 49} 705--710
  \urlprefix\url{http://dx.doi.org/10.1051/jphys:01988004905070500}

\bibitem{Arunchander2015}
Arunchander A, Peera S~G, Parthiban V, Akula S, Kottakkat T, Bhat
S~D and Sahu
  A~K 2015 {\em {RSC} Adv.\/} {\bf 5} 75218--75228
  \urlprefix\url{http://dx.doi.org/10.1039/C5RA15233J}

\bibitem{Anderson2016}
Anderson W~C, Rhinehart J~L, Tennyson A~G and Long B~K 2016 {\em J.
Am. Chem.
  Soc.\/} {\bf 138} 774--777
  \urlprefix\url{http://dx.doi.org/10.1021/jacs.5b12322}

\bibitem{Striegel2009}
Striegel A~M, Yau W~W, Kirkland J~J and Bly D~D 2009 {\em Modern
Size-Exclusion
  Liquid Chromatography\/} (Wiley-Blackwell)
  \urlprefix\url{http://dx.doi.org/10.1002/9780470442876}

\bibitem{Kuhn1936}
Kuhn W 1936 {\em Angewandte Chemie\/} {\bf 49} 858--862
  \urlprefix\url{http://dx.doi.org/10.1002/ange.19360494803}

\bibitem{Guida1998}
Guida R and Zinn-Justin J 1998 {\em J. Phys. A: Math. Gen.\/} {\bf
31}
  8103--8121 \urlprefix\url{http://dx.doi.org/10.1088/0305-4470/31/40/006}

\bibitem{ZinnJustin2002}
Zinn-Justin J 2002 {\em Quantum Field Theory and Critical
Phenomena\/} (Oxford
  University Press ({OUP}))
  \urlprefix\url{http://dx.doi.org/10.1093/acprof:oso/9780198509233.001.0001}

\bibitem{Aronovitz1986}
Aronovitz J and Nelson D 1986 {\em J. Phys. France\/} {\bf 47}
1445--1456
  \urlprefix\url{http://dx.doi.org/10.1051/jphys:019860047090144500}

\bibitem{Cannon1991}
Cannon J~W, Aronovitz J~A and Goldbart P 1991 {\em J. Phys. I
France\/} {\bf 1}
  629--645 \urlprefix\url{http://dx.doi.org/10.1051/jp1:1991159}

\bibitem{Jagodzinski1992}
Jagodzinski O, Eisenriegler E and Kremer K 1992 {\em J. Phys. I
France\/} {\bf
  2} 2243--2279 \urlprefix\url{http://dx.doi.org/10.1051/jp1:1992279}

\bibitem{Bishop1988}
Bishop M and Saltiel C~J 1988 {\em The Journal of Chemical
Physics\/} {\bf 88}
  6594 \urlprefix\url{http://dx.doi.org/10.1063/1.454446}

\bibitem{Benhamou1985}
Benhamou M and Mahoux G 1985 {\em Journal de Physique Lettres\/}
{\bf 46}
  689--693
  \urlprefix\url{http://dx.doi.org/10.1051/jphyslet:019850046015068900}

\bibitem{Diehl1989}
Diehl H~W and Eisenriegler E 1989 {\em J. Phys. A: Math. Gen.\/}
{\bf 22}
  L87--L91 \urlprefix\url{http://dx.doi.org/10.1088/0305-4470/22/3/005}

\bibitem{Blavatska2011}
Blavatska V, von Ferber C and Holovatch Y 2011 {\em Condensed Matter
Physics\/}
  {\bf 14} 33701 \urlprefix\url{http://dx.doi.org/10.5488/CMP.14.33701}

\bibitem{Zifferer1999a}
Zifferer G 1999 {\em The Journal of Chemical Physics\/} {\bf 110}
4668--4677
  \urlprefix\url{http://dx.doi.org/10.1063/1.478350}

\bibitem{Zifferer1999b}
Zifferer G 1999 {\em Macromolecular Theory and Simulations\/} {\bf
8} 433--462
  \urlprefix\url{http://dx.doi.org/10.1002/(SICI)1521-3919(19990901)8:5<433::AID-MATS433>3.0.CO;2-C}

\bibitem{Victor1990}
Victor J~M and Lhuillier D 1990 {\em The Journal of Chemical
Physics\/} {\bf
  92} 1362--1364 \urlprefix\url{http://dx.doi.org/10.1063/1.458147}

\bibitem{Hoogerbrugge1992}
Hoogerbrugge P~J and Koelman J~M~V~A 1992 {\em Europhysics Letters
({EPL})\/}
  {\bf 19} 155--160
  \urlprefix\url{http://dx.doi.org/10.1209/0295-5075/19/3/001}

\bibitem{Espanol1995}
Espa{\~{n}}ol P and Warren P 1995 {\em Europhysics Letters
({EPL})\/} {\bf 30}
  191--196 \urlprefix\url{http://dx.doi.org/10.1209/0295-5075/30/4/001}

\bibitem{Schlijper1995}
Schlijper A~G 1995 {\em Journal of Rheology\/} {\bf 39} 567
  \urlprefix\url{http://dx.doi.org/10.1122/1.550713}

\bibitem{Kong1997}
Kong Y, Manke C~W, Madden W~G and Schlijper A~G 1997 {\em The
Journal of
  Chemical Physics\/} {\bf 107} 592
  \urlprefix\url{http://dx.doi.org/10.1063/1.474420}

\bibitem{Spenley2000}
Spenley N~A 2000 {\em Europhysics Letters ({EPL})\/} {\bf 49}
534--540
  \urlprefix\url{http://dx.doi.org/10.1209/epl/i2000-00183-2}

\bibitem{Symeonidis2005}
Symeonidis V, Karniadakis G~E and Caswell B 2005 {\em Phys. Rev.
Lett.\/} {\bf
  95} \urlprefix\url{http://dx.doi.org/10.1103/PhysRevLett.95.076001}

\bibitem{Jiang2007}
Jiang W, Huang J, Wang Y and Laradji M 2007 {\em The Journal of
Chemical
  Physics\/} {\bf 126} 044901
  \urlprefix\url{http://dx.doi.org/10.1063/1.2428307}

\bibitem{Nardai2009}
Nardai M~M and Zifferer G 2009 {\em The Journal of Chemical
Physics\/} {\bf
  131} 124903 \urlprefix\url{http://dx.doi.org/10.1063/1.3231854}

\bibitem{Ilnytskyi2007}
Ilnytskyi J~M and Holovatch Y 2007 {\em Condensed Matter Physics\/}
{\bf 10}
  539 \urlprefix\url{http://dx.doi.org/10.5488/CMP.10.4.539}

\bibitem{Ilnytskyi2008}
Ilnytskyi J~M, Patsahan T, Holovko M, Krouskop P~E and Makowski M~P
2008 {\em
  Macromolecules\/} {\bf 41} 9904--9913
  \urlprefix\url{http://dx.doi.org/10.1021/ma801045z}

\bibitem{Ilnytskyi2011}
Ilnytskyi J~M, Patsahan T and Soko{\l}owski S 2011 {\em The Journal
of Chemical
  Physics\/} {\bf 134} 204903
  \urlprefix\url{http://dx.doi.org/10.1063/1.3592562}

\bibitem{Ilnytskyi2013}
Ilnytskyi J~M, Sokolowski S and Patsahan T 2013 {\em Condensed
Matter
  Physics\/} {\bf 16} 13606
  \urlprefix\url{http://dx.doi.org/10.5488/CMP.16.13606}

\bibitem{Ilnytskyi2016}
Ilnytskyi J~M, Bryk P and Patrykiejew A 2016 {\em Condensed Matter
Physics\/}
  {\bf 19} 13609 \urlprefix\url{http://dx.doi.org/10.5488/CMP.19.13609}

\bibitem{Groot1997}
Groot R~D and Warren P~B 1997 {\em The Journal of Chemical
Physics\/} {\bf 107}
  4423 \urlprefix\url{http://dx.doi.org/10.1063/1.474784}

\bibitem{Solc1971a}
\v{S}olc K 1971 {\em The Journal of Chemical Physics\/} {\bf 54}
2756
  \urlprefix\url{http://dx.doi.org/10.1063/1.1675241}

\bibitem{Solc1971b}
\v{S}olc K and Stockmayer W~H 1971 {\em The Journal of Chemical
Physics\/} {\bf
  55} 335 \urlprefix\url{http://dx.doi.org/10.1063/1.1675527}

\bibitem{Sciutto1996}
Sciutto S~J 1996 {\em J. Phys. A: Math. Gen.\/} {\bf 29} 5455--5473
  \urlprefix\url{http://dx.doi.org/10.1088/0305-4470/29/17/019}

\bibitem{Sciutto1994}
Sciutto S~J 1994 {\em J. Phys. A: Math. Gen.\/} {\bf 27} 7015--7034
  \urlprefix\url{http://dx.doi.org/10.1088/0305-4470/27/21/017}

\bibitem{Rudnick1986}
Rudnick J and Gaspari G 1986 {\em J. Phys. A: Math. Gen.\/} {\bf 19}
L191--L193
  \urlprefix\url{http://dx.doi.org/10.1088/0305-4470/19/4/004}

\bibitem{Gaspari1987}
Gaspari G, Rudnick J and Beldjenna A 1987 {\em J. Phys. A: Math.
Gen.\/} {\bf
  20} 3393--3414 \urlprefix\url{http://dx.doi.org/10.1088/0305-4470/20/11/041}

\bibitem{Wei1990}
Wei G and Eichinger B~E 1990 {\em Macromolecules\/} {\bf 23}
4845--4855
  \urlprefix\url{http://dx.doi.org/10.1021/ma00224a013}

\bibitem{Sciutto1995}
Sciutto S~J 1995 {\em J. Phys. A: Math. Gen.\/} {\bf 28} 3667--3679
  \urlprefix\url{http://dx.doi.org/10.1088/0305-4470/28/13/012}

\end{thebibliography}
\end{document}